\documentclass[10pt,twoside,onecolumn]{IEEEtran}
\input{epsf}
\usepackage{framed}
\usepackage{color}

\definecolor{bgrd}{rgb}{1,1,1}
\definecolor{grey}{rgb}{0.9,0.9,0.6}
\definecolor{gray}{rgb}{0.5,0.5,0.5}
\definecolor{dkr}{rgb}{0.6,0.2,0.2}
\definecolor{dkg}{rgb}{0,0.5,0}
\definecolor{dkb}{rgb}{0.0,0.1,0.7}
\definecolor{light-gray}{gray}{0.85}

\newcommand{\test}{\mbox{$\begin{array}{c}
\stackrel{\stackrel{\textstyle {\cal H}_1}{\textstyle >}}
{\stackrel{\textstyle <}{\textstyle {\cal H}_0}} \end{array}$}}

\usepackage{theorem}

\renewcommand{\P}{\mathbb{P}}

\newcommand{\beq}{\begin{equation}}
\newcommand{\eeq}{\end{equation}}
\newcommand{\beqa}{\begin{eqnarray}}
\newcommand{\eeqa}{\end{eqnarray}}
\newcommand{\dfz}{\triangleq}

\newcommand{\bd}{{\mathbf{d}}}

\newcommand{\bu}{{\mathbf{u}}}
\newcommand{\bz}{{\mathbf{z}}}
\newcommand{\bn}{{\mathbf{n}}}

\newcommand{\by}{{\mathbf{y}}}

\newcommand{\bw}{{\mathbf{w}}}

\newcommand{\bx}{{\mathbf{x}}}

\newcommand{\bi}{{\mathbf{i}}}

\newcommand{\VAR}{\textnormal{$\mathbb{V}$}}

\newcommand{\E}{\mathbb{E}}

\newcommand{\cN}{{\cal N}}

\newcommand{\cD}{{\cal D}}
\newcommand{\cH}{{\cal H}}

\usepackage{cite}
\usepackage{multirow}
\usepackage{rotating}
\usepackage{hhline}
\usepackage{amssymb}
\usepackage{mathrsfs}
\usepackage{graphicx}
\usepackage{epstopdf}
\usepackage{amsmath}
\usepackage{empheq}

\usepackage{bm}

\usepackage{soul}
\usepackage[vlined,ruled]{algorithm2e}

\newcommand{\sign}{\mbox{sign}}

\usepackage{xcolor, framed}

\begin{document}


\title{Decision-Making Algorithms for Learning and Adaptation with Application to COVID-19 Data}
\author{Stefano~Marano, \IEEEmembership{Senior Member, IEEE} and Ali H. Sayed, \IEEEmembership{Fellow, IEEE} 
\thanks{S.~Marano is with DIEM, University of Salerno, via Giovanni Paolo~II 132, I-84084, Fisciano (SA), Italy (e-mail: marano@unisa.it). A.~H.~Sayed is with the Ecole Polytechnique Federale de Lausanne EPFL, School of Engineering, CH-1015 Lausanne, Switzerland (e-mail: ali.sayed@epfl.ch).}
} 

\maketitle

\begin{abstract}

This work focuses on the development of a new family of decision-making algorithms for adaptation and learning, which are specifically tailored to decision problems and are constructed by building up on first principles from decision theory. A key observation is that estimation and decision problems are structurally different and, therefore, algorithms that have proven successful for the former need not perform well when adjusted for decision problems. We propose a new scheme, referred to as BLLR (barrier log-likelihood ratio algorithm) and demonstrate its applicability to real-data from the COVID-19 pandemic in Italy. The results illustrate the ability of the design tool to track the different phases of the outbreak.
\end{abstract}


\section{Introduction}
Performing inference by a network of interconnected agents is the primary goal in many practical applications, as seen in~\cite{chamberland,PreddKulkarniPoor-magazine06,akyildiz-survey,Varshney:book,viswanathan97,blum97,tsitsiklis93,Tsitsiklis88,SENMA,doasplet05,spawc06,tong:C-SENMA-LDPC}.
In these works, the desired inference is typically obtained at a central unit that processes the data received from remote agents. Substantial improvements in system robustness, reliability and scalability can be obtained with fully-flat architectures without a central unit at the cost of more capable agents. These agents will now be required to obtain local inference solutions in a fully distributed manner by exploiting data exchanges among neighboring agents. 

In fully-flat architectures the signal-processing mechanism can be based on consensus strategies~\cite{boyd-infocom,running-cons,asymptotic-rc,kar-moura-stsp,mouraetal2011,mouraetal2012,mouraetal2012_2} or diffusion strategies~\cite{CattivelliSayedDetection,SayedSPmag,SayedNOW2014,SayedprocIEEE,chen-sayed-IT1,chen-sayed-IT2,TowficChenSayedIT2016,BracaetalIT,MattaSIPN16,MaranoSayedIT19,ICASSP2020,maranosayedsubmitted2020,ChenJSTSP17}. Among the latter class, the most successful strategy is the ATC (adapt-then-combine) fusion rule, which consists of two steps. First, in the adaptation step, after collecting a new observation about the phenomenon of interest,  the agent updates its state (inference statistic) by incorporating the information provided by the fresh measurement. Then, the updated state is combined
with those of nearby agents and progressively diffused throughout the network. The adaptation step is based on the popular LMS (least-mean-square) algorithm, see, e.g.,~\cite{Sayed2008adaptive}, while the second step consists of computing a convex combination of the states of neighboring agents.

Historically, in learning and adaptation contexts, \emph{estimation}
problems were considered first, which led to the adoption of the LMS algorithm for the adaptation step, because of its well-known adaptation properties~\cite{Sayed2008adaptive,SayedSPmag,SayedNOW2014}. When dealing with \emph{decision} problems, it appears natural to maintain the LMS protocol as the basic engine due to its simplicity in order to track drifts in the state of nature. For this reason, most works in the literature addressing distributed decision problems exploit the ATC diffusion strategy, in the form of LMS iterates followed by a convex combination of states, as originally designed for estimation problems~\cite{BracaetalIT,MattaSIPN16,MaranoSayedIT19,ICASSP2020,maranosayedsubmitted2020}.

\subsection{Contribution and Scope}
In this paper we design
alternative decision-making algorithms that are specifically tailored to decision problems, by building up on first principles from decision theory rather than relying directly on the LMS update. 
We introduce performance indexes that quantify the tradeoff between learning and adaptation. Using these performance figures, we show that the proposed agents' updating rule outperforms the one based on the LMS iteration. 
In this contribution, no network aspects are considered and the focus is on the operation of a single agent. In addition, we limit our study to the case of two possible states of nature, say $\cH_0$ and $\cH_1$, which are known to the decision maker. This means that at any time epoch,
the observations collected by the agent are independently drawn from one of two 
distributions, but we do not know which, and this underlying distribution is allowed to change at any time according to arbitrary patterns of the kind $\dots \cH_0 \mapsto \cH_1 \mapsto \cH_0 \mapsto 
\cH_1 \dots $. Extensions to more than two states of nature are left for future studies. 

We illustrate that the learning and adaptation technique developed in this paper is useful in tracking different phases of the COVID-19 pandemic. As an example, the proposed tool is demonstrated on pandemic data from Italy.

\subsection{Notation}

Boldface symbols denote random variables and normal font their realizations and deterministic quantities.
For scalars, the time index (or algorithm iteration number) is enclosed in parentheses. 
Thus, for instance, $\bx(n)$ denotes the random scalar $\bx$ at time~$n$. 
Conversely, in the case of vectors, the time dependence is indicated by a subscript, as, for example, $\bu_i$ denotes a random vector $\bu$ evaluated at time $i$. Superscript $T$ denotes vector transposition.
Statistical expectation, variance, and probability operators are denoted by $\E$, $\VAR$, and $\P$, respectively. They always are computed under the hypothesis in force $\cH_h$, and the pertinent subscript $h=0,1$ is usually added to the operator symbol.

The remaining part of this article is organized as follows. Section~\ref{sec:genesis} discusses the genesis of LMS in decision contexts. The proposed alternative to LMS is presented in Sec.~\ref{sec:proposed} and its performance is investigated in Sec.~\ref{sec:per} in terms of the criteria discussed in Sec.~\ref{sec:perfass}. Examples using synthetic data are given in Sec.~\ref{sec:examples} while an application to COVID-19 pandemic time-series is discussed in Sec.~\ref{sec:covid}. Section~\ref{sec:conclusion} contains conclusive remarks.

\section{Genesis of the LMS Algorithm in Decision Contexts}
\label{sec:genesis}

\label{sec:SALMSest}

Let us start by considering an \emph{estimation} problem. Let $\bd\in \Re$ be a zero-mean scalar random variable with variance $\E\bd^2>0$, and $\bu \in \Re^{M}$ a zero-mean random vector with positive-definite covariance matrix $\E \bu \bu^T >0$. The quantity $\bd$ is unknown while $\bu$ is observed.
The goal is to solve the optimization problem $\min_w J(w)$,
where $w \in \Re^{M}$ is a weight vector and $J(w) : \Re^{M} \mapsto \Re$ represents a cost function that quantifies the penalty incurred when the unknown $\bd$ 
is replaced by the linear transformation $\bu^T w$ of the observation.
One common choice is the quadratic cost function $J(w)=\E(\bd -\bu^T w)^2$, in which case the solution $w^{\rm o}$ 
is given by $w^{\rm o}=(\E \bu \bu^T)^{-1} \E \bd\bu$,
and the \emph{linear least-mean-square estimator} of $\bd$ given $\bu$ is $\widehat \bd=\bu^T w^{\rm o}$~\cite[Th.~8.1, p.~142]{Sayed2008adaptive}.

A recursive solution to the optimization problem $\min_w J(w)$ 
with quadratic cost function is provided by the \emph{steepest-descent} algorithm: set $w_{0}$ equal to some initialization vector, and iterate as follows:
\begin{align}
w_i=w_{i-1} + \mu \big [\E \bd\bu- \E \bu \bu^T \,  w_{i-1} \big ], \quad i=1,2,\dots,
\label{eq:steepest}
\end{align}
where the step-size $\mu>0$ is sufficiently small (less than 2 divided by the largest eigenvalue of matrix $\E \bu \bu^T$), see~\cite[Th.~8.2, p.~147]{Sayed2008adaptive}.
It can be shown that $\E \bd\bu- \E \bu \bu^T \,  w_{i-1} =-\nabla J(w_{i-1})$, which makes it possible to rewrite~(\ref{eq:steepest}) in terms of the gradient vector $\nabla J(w_{i-1})$. The resulting expression is useful when alternative cost functions are used.

What is especially relevant in the adaptive framework is the consideration that the quantities $\E \bu^T \bu$ and $\E \bd\bu$ may not be known beforehand and are expected to vary over time. 
In these situations, assuming that we have access to streaming data in the form of a sequence of realizations $\{d(i), u_i\}_{i\ge 1}$ of $\bd$ and $\bu$, a viable alternative to~(\ref{eq:steepest}) is obtained if we drop the expectation signs and replace the random variables by their current realizations, yielding the following algorithm: set $w_{0}=$ some initial guess, 
\begin{align}
w_i=w_{i-1} + \mu u_i \big [d(i)-  u_i^T \,  w_{i-1} \big ], \quad i=1,2,\dots,
\label{eq:LMS0}
\end{align}
with a sufficiently small $\mu$.
This \emph{stochastic gradient} approximation (because the true gradient is replaced by a \emph{noisy} version thereof) is known as the LMS algorithm, see~\cite[Th.~10.1, p.~166]{Sayed2008adaptive}.
The LMS algorithm learns the data statistics and at the same time is able to track statistical drifts, which are essential characteristics for the design of cognitive intelligent inference systems with learning and adaptation properties. 

We now move from an estimation to a decision context, paralleling the way in which this happened in the literature.
Suppose $M=1$, namely ${w_i}={w(i)}$ and ${u_i}={u(i)}$ are scalars, and suppose also $u(i)=1$ for all $i$. 
By assuming independent and identically distributed (IID) data $\{\bd(i)\}_{i\ge1}$, formal substitution in~(\ref{eq:LMS0})
gives: $\bw(0)=0$, 
\begin{align}
\bw(i)&=\bw(i-1)+ \mu  [ \bd(i)- \bw(i-1)], \qquad i\ge1, \label{eq:LMSdet}
\end{align}
Note that the right-hand side of~(\ref{eq:LMSdet}) is a convex combination: $\mu \bd(i) + (1-\mu) \bw(i-1)$.
Iterating~(\ref{eq:LMSdet}), we get the output of the LMS algorithm in the form:
\begin{align}
\bw(i) = \sum_{k=0}^{i-1} \mu (1-\mu)^k \bd(i-k),
\label{eq:LMSexpl}
\end{align}
and we have
\begin{subequations}
\begin{align}
\E  \bw(i) &=  \, [1-(1-\mu)^i] \, \E \bd,  \\
\VAR \bw(i) & = [1-(1-\mu)^{2i}] \frac{\mu}{2-\mu} \, \VAR \bd.
\end{align}
\label{eq:EVAR}%
\end{subequations}
From~(\ref{eq:EVAR}), we see that the output of the algorithm approximates $\E \bd$ when the number $i$ of iterations is sufficiently large and 
the step-size $\mu$ is $\ll 1$. This property, along with the inherent adaptation ability, motivated the use of~(\ref{eq:LMSdet}) in \emph{decision} problems. Indeed, the algorithm formalized in~(\ref{eq:LMSdet}) represents the basic building block for the development of 
adaptation and learning diffusion algorithms over networks faced with decision problems, which has been addressed in a series of 
papers~\cite{ICASSP2020,maranosayedsubmitted2020,MaranoSayedIT19,BracaetalIT,MattaSIPN16}.

\subsection{Alternative Derivation} 
\label{sec:alternative}

Since the main motivation for this paper is to explore alternatives to the LMS block, it is important to consider alternative contexts in which LMS arises 
and analyze the related motivation for its usage. Two contexts in which the LMS is used are now briefly discussed. The first is a minimax formulation of an estimation problem, and the second is linked to decision procedures. 

An algorithm similar to that shown in~(\ref{eq:LMS0}) emerges in the following scenario, see~\cite[p.~731]{Sayed2008adaptive}.
Suppose that we observe $\{d(i), u_i\}_{i\ge 1}$, modeled as $d(i)=s(i)+v(i)$, with $s(i)\dfz u_i^Tw^{\rm o}$, for some 
sequence of vectors $\{u_i\}_{i\ge1}$ satisfying a ``persistence of excitation''
condition $\sum_{i=1}^\infty u_i u_i^T=\infty$, and some unknown $w^{\rm o}\in \Re^{M \times 1}$. Here $v(i)\in \Re$ is a ``noise'' term with finite energy $\sum_{i=1}^\infty v^2(i)<\infty$. Let $\widehat s(i|i-1)$ denote a strictly causal estimator of $s(i)$ based on the data $\{d(k)\}_{k=0}^{i-1}$.
Then, the LMS algorithm is the optimal solution to the min-max 
problem~\cite{Hassibi-Sayed-Kailath}:
\begin{align}
\inf_{\{\widehat s(i|i-1)\}} \sup_{\{v(i)\},w^{\rm o}} \: \frac{\sum_{i=1}^\infty [\widehat s(i|i-1)-s(i)]^2}{\mu^{-1} \| w^{\rm o} \|^2+ \sum_{i=1}^\infty v^2(i)},
\label{eq:min-max}
\end{align}
and, moreover, the fraction in~(\ref{eq:min-max}) takes value 1 at optimality. 
In particular, let $w_{0}=0$ and, for $1 \le i < n$, consider the iteration
\begin{subequations}
\begin{align} 
\widehat s(i|i-1)&=u_i^T w_{i-1}, \label{eq:LMSa}\\
 w_i&=w_{i-1}+ \mu u_i [ d(i)- u_i^T w_{i-1}]. \label{eq:LMSb}
\end{align}
\label{eq:LMS}%
\end{subequations}
It can be shown that this version of the LMS algorithm satisfies the 
following (robustness) condition for every $i<n$:
\begin{align}
\frac{\sum_{k=1}^i [\widehat s(k|k-1)-s(k)]^2}{\mu^{-1} \| w^{\rm o} \|^2+ \sum_{k=1}^{i-1} v^2(k)} < 1, 
\label{eq:min-max-finite}
\end{align}
if, and only if, all the matrices $\{ \mu^{-1} I - u_i^T u_i\}_{i=0}^{n-1}$ are positive-definite~\cite[Alg.~45.4, p.~731]{Sayed2008adaptive}.

It is also useful to mention that the LMS has been advocated in \emph{decision} problems and, specifically, in the context of continuous inspection schemes and related control charts. As seen in~\eqref{eq:LMSexpl}, LMS employs exponentially-scaled weights, which is exactly the idea behind  
the geometric moving average control charts, see~\cite[Sec.\ 2.1.2, p.\ 28]{basseville-book} and~\cite[Sec.\ 8.1.2, p.\ 373]{tartakovsky-book}.
In these contexts, LMS is known under the name of GMA (Geometric Moving Average) or EWMA (Exponentially Weighted Moving Average).
We refer to~\cite{basseville-book,tartakovsky-book} for details.

\section{Proposed Algorithm: Barrier LLR}
\label{sec:proposed}

As discussed in the previous section, in the literature of adaptation and learning, decision problems have been approached by exploiting schemes and protocols initially conceived for estimation problems. Since decision and estimation problems are structurally different in many respects, 
it makes sense to start anew, with the goal of exploring possible alternatives to the LMS component with better performance for decision tasks.
The idea is to modify a classical decision algorithm in order to make it more suitable to adaptation contexts.
Let us consider a standard binary decision problem in which IID data $\{\bx(i)\}_{i\ge1}$ are observed and the following binary hypothesis test must be solved:
\begin{align}
\label{eq:test}
\begin{array}{ll}
\cH_1: & \bx(i) \sim f_1(x), \quad i=1,2,\dots \\
\cH_0: & \bx(i) \sim f_0(x), \quad i=1,2,\dots 
\end{array}
\end{align}
where $f_{1,0}(x)$ are the probability density functions (PDFs) of the data under the two hypotheses $\cH_1$ and $\cH_0$, respectively. These PDFs are assumed to exist and are known to the decision maker.
It is well-known that under the most popular optimality criteria, the optimal decision maker exploits the log-likelihood of the data~\cite{poorbook}, which is
\begin{align} \label{eq:LL}
\bd(i)=\log \frac{f_1(\bx(i))}{f_0(\bx(i))}.
\end{align}
Exploiting the IID property of the observations, the optimal decision based on vector $[\bx(1),\dots,\bx(n)]^T$ is
\begin{align}
\bz(n)=\sum_{i=1}^n \bd(i) = \sum_{i=1}^n \log \frac{f_1(\bx(i))}{f_0(\bx(i))} \test \gamma,
\label{eq:z}
\end{align}
where $\gamma$ is a suitable threshold, chosen according to the desired optimality criterion~\cite{poorbook}.
Expression~\eqref{eq:z} can be regarded as a random walk: $\bz(0)=0$,
\begin{align}
\bz(i) = \bz(i-1) + \bd(i), \qquad i \ge 1,
\label{eq:walkz}
\end{align}
with step $\bd(i)$. 
The following relationships are well known [the argument ``$(i)$'' is suppressed when non-essential]:
\begin{align}
\label{eq:twodiv}
\E_1[ \bd]=D_{10} >0, \qquad \E_0[ \bd]=-D_{01} <0
\end{align} 
where $D_{10}= \int f_1(x) \log [f_1(x)/f_0(x)] \, dx$ and  $D_{01}= \int f_0(x) \log [f_0(x)/f_1(x)] \, dx$
denote the two Kullback-Leibler (KL) distances (or divergences) between the PDFs $f_1(x)$ and $f_0(x)$~\cite{CT2}.
In~\eqref{eq:twodiv} we have assumed that that these KL distances exist and are strictly positive, which implies that $f_1(x)$ and $f_0(x)$ are distinct over a set of nonzero probability. 

\vspace*{3pt}\emph{\textsc{\textbf{Assumptions.}}} The following assumptions are used throughout the paper. Under $\cH_h$, $h=0,1$, the random variables $\{\bd(i)\}_{i\ge1}$ are continuous, with finite first- and second-order moments, and their probability distribution admits a density with respect to the usual Lebesgue measure. In addition, 
$\E_1 \bd > 0$, $\E_0 \bd < 0$, $\P_h(\bd>0)>0$ and $\P_h(\bd<0)>0$, $h=0,1$.
\hfill$\square$

\vspace*{3pt} 
In \eqref{eq:z}, we see that the optimal decision maker compares to a threshold the value of a random walk with positive drift under $\cH_1$ and negative drift under $\cH_0$.
This optimal \emph{learning} scheme is not \emph{adaptive}, as is easily revealed by the following informal arguments.
Suppose that the state of nature is $\cH_1$ for time steps $1 \le i \le n$, and then switches to $\cH_0$. 
Observations are IID under each hypothesis, but their distribution is different under the two hypotheses. 
Assuming $n \gg1$, with high probability the decision statistic $\bz(n)$ takes on very large values because the random walk is drifting to $+\infty$ for $1 \le i \le n$. For $i>n$ the drift is negative but the random walk ``starts'' at $\bz(n)$, implying that the time required to approach the negative values that are typical of hypothesis $\cH_0$ is very large. 
A straightforward way to prevent $\{\bz(i)\}_{i\ge0}$ from reaching extreme values is to introduce two barriers $-a<0<b$, as follows:
$\bz(0)=0$ and for $i \ge 1$:
\begin{align}
\bz(i)=
\begin{cases} 
-a,  & \bz(i-1) + \bd(i) \le - a, \\
\bz(i-1) + \bd(i), & -a < \bz(i-1) + \bd(i) < b, \\
b,  & \bz(i-1) + \bd(i)  \ge b .
\end{cases}
\label{eq:BLLR}
\end{align}
A more compact expression for the iteration in~\eqref{eq:BLLR} is: $\bz(0)=0$ and
\begin{align}
\bz(i)=\inf \big \{ b, \sup \{-a,\bz(i-1)+ \bd(i) \} \big \}, \quad i \ge 1.
\label{eq:BLLRcompact}
\end{align}

The lower and upper barriers limit the range of values that $|\bz(i)|$ takes on, and hence we can tradeoff adaptation and learning by a careful choice of $a$ and $b$.
Large values favor the learning (decision) ability of the system, while small values favor its adaptation ability. 
In the following, the decision procedure based on comparing~\eqref{eq:BLLR} to a threshold $\gamma \in (-a,b)$ will be referred to as the \emph{barrier log-likelihood ratio} (BLLR) test:
the BLLR decision at any arbitrary time epoch $n$ is
\begin{align}
\bz(n) \test \gamma.
\label{eq:plaintest}
\end{align}
The BLLR decision procedure~\eqref{eq:plaintest}, with $\bz(n)$ shown in~\eqref{eq:BLLR} or~\eqref{eq:BLLRcompact}, represents the proposed alternative. 
For easy reference, the BLLR and LMS procedures are summarized in Algorithms~\ref{alg:BLLR} and~\ref{alg:LMS}.

\begin{algorithm}[t]
\small{
\DontPrintSemicolon
\KwIn{input sequence $\{d(n)\}_{n=1}^{n_{\max}}$; initialization $z(0)$; barriers $a$, $b$}  
\KwOut{decision statistic $\{z(n)\}_{n=1}^{n_{\max}}$} 
\vspace*{5pt}
$n=0$; \\
\textbf{while} $n<n_{\max}$ \\
$\quad n=n+1$ \\
$\quad z(n)= \inf \{ b, \sup \{ -a, z(n-1) + d(n)\} \}$ \\
\textbf{end}
\caption{\textbf{BLLR}
\label{alg:BLLR}}
}
\end{algorithm}
\begin{algorithm}[t]
\small{
\DontPrintSemicolon
\KwIn{input sequence $\{d(n)\}_{n=1}^{n_{\max}}$; initialization $w(0)$; step-size $\mu$}  
\KwOut{decision statistic $\{w(n)\}_{n=1}^{n_{\max}}$} 
\vspace*{5pt}
$n=0$; \\
\textbf{while} $n<n_{\max}$ \\
$\quad n=n+1$ \\
$\quad w(n)= \mu d(n)+(1-\mu) w(n-1)$ \\
\textbf{end}
\caption{\textbf{LMS}}
\label{alg:LMS}}
\end{algorithm}
Typical choices for the threshold appearing in~\eqref{eq:plaintest} are: $\gamma=0$, which, in the unbounded case of $a,b\to \infty$ corresponds to the maximum likelihood (ML) decision criterion
adopted, among other cases, in the Bayesian framework in which the two hypotheses are equally likely; the mid-point threshold $\gamma=\frac{b-a}{2}$; or the value of $\gamma$
for which the error probability of deciding $\cH_1$ under state of nature $\cH_0$ takes on a prescribed value (false alarm criterion), as in the Neyman-Pearson formulation~\cite{poorbook}. Likewise, when considering the LMS iterate~\eqref{eq:LMSdet}, a test similar to~\eqref{eq:plaintest} is used and the threshold is chosen with the same criteria. For LMS, the mid-point threshold is $\gamma=(D_{10}-D_{01})/2$.

Suppose we know that $\cH_0$ is in force for $1\le i \le n$ and $\cH_1$ is in force for all $i > n$, with IID data under each hypothesis, and the change epoch $n$ is unknown. The celebrated Page's test for quickest detection of a change in the state of nature is obtained from~\eqref{eq:BLLR} by setting $a=0$ and letting $b$ be the decision threshold: once the decision statistic hits the value~$b$, the change in the state of nature $\cH_0 \mapsto \cH_1$  is declared and the test stops~\cite{basseville-book}. 
This reveals that BLLR test~\eqref{eq:BLLR} is a generalization of Page's test. Indeed, the BLLR test in~\eqref{eq:BLLR} can be seen as an infinite sequence of Page's tests for successively detecting the changes $\cH_0 \mapsto \cH_1 \mapsto \cH_0 \mapsto \cH_1 \dots$. 
To see this, let us assume that $\cH_0$ is true and set $\bz(0)=-a$. The BLLR test is equivalent to a Page's test with threshold $a+b$ for detecting the change  $\cH_0 \mapsto \cH_1$, followed by a \emph{sign-reversed} Page's test initialized at $b$, driven by negative drifts, with threshold $a+b$, to detect the successive change  $\cH_1 \mapsto \cH_0$, and so forth indefinitely.\footnote{We are making the simplifying assumption that the hits at the thresholds are really due to change in the state of nature, not to error events.} 
In turn, Page's test can be regarded as a sequence of Wald's SPRTs (sequential probability ratio tests)~\cite{wald,wald-wolfowitz-AMS}, which reveals that the decision algorithm shown in~\eqref{eq:BLLR} and~\eqref{eq:plaintest} is a modified version of a sequence of SPRTs. Not surprisingly, the performance analysis of BLLR relies on standard results of sequential analysis, some results of which are collected in Appendices~\ref{app:exact}-\ref{app:Fredholm}, for self-consistency.

In view of the analogy with sequential analysis, our approach is close in spirit to the SPRT approach pursued in~\cite{Sayed-CooperativeSensing} for cooperative sensing. As done in~\cite{Sayed-CooperativeSensing}, in Sec.~\ref{sec:covid} we resort to the GLRT (generalized likelihood ratio test) approach to deal with the presence of unknown parameters.
However, the nature of these parameters and the corresponding estimates are structurally different from those in~\cite{Sayed-CooperativeSensing}, resulting in substantially different decision procedures.

\section{Performance Assessment}
\label{sec:perfass}

\subsection{Performance Criteria} 

\subsubsection{Performance for BLLR Test}
\label{sec:PC}

We introduce two performance indexes:  the \emph{error rate} $r$, related to the learning capability, and the \emph{expected delay} $\Delta$ that quantifies the adaptation capability. 

Let us consider the learning aspects first. Perhaps, the most natural performance figures would be the probability that $\lim_{i\to\infty} \bz(i)$ exceeds $\gamma$ under $\cH_0$, and
the probability that $\lim_{i\to\infty} \bz(i)$ goes below $\gamma$ under $\cH_1$. In general, these steady-state probabilities are guaranteed to exist~\cite{afanaseva,borovkov}, however they are not easy to compute and do not lead to simple closed-form expressions from which insights can be easily gained. We instead introduce performance figures whose computation is tractable. 
For $h=0,1$, consider the following quantity, defined with the state of nature $\cH_h$ held fixed:\footnote{In the following we also use the qualification ``steady-state'' state of nature to signify that the state of nature is assumed forever constant.}
\begin{align} \label{eq:Tgeneric}
T_h(z_0;z_1)= \E_h \inf_{i\ge1} \{i: \bz(i) \gtreqless z_1 ; \textnormal{ with } \bz(0)=z_0\},
\end{align}
wherein the sign $\ge$ applies if $z_0 < z_1$, and $\le$ applies if $z_0 > z_1$. The quantity in~\eqref{eq:Tgeneric} represents the expected time
to reach the value $z_1$ starting from $z_0$, under hypothesis $\cH_h$.

Using~\eqref{eq:Tgeneric}, the learning ability of the system is quantified by the two indexes
\begin{align} \label{eq:Tspecificrate}
T_0(-a;\gamma) \qquad \textnormal{and} \qquad T_1(b;\gamma).
\end{align}
The interpretation is as follows. The quantity $T_0(-a;\gamma)$ represents the expected time to cross the threshold $\gamma$, yielding a decision in favor of $\cH_1$, in the $\cH_0$ steady-state situation, when the BLLR iteration is initialized to $\bz(0)=-a$, which
we call the ``typical'' value taken by the statistic under $\cH_0$. Likewise, $T_1(b;\gamma)$ represents  the expected time to cross the threshold yielding the $\cH_0$ decision, in the $\cH_1$ steady-state situation, when the BLLR iteration is initialized at the ``typical'' value under $\cH_1$, which is $\bz(0)=b$.
Note that these quantities are related to --- but different from --- the expected time between false alarms and miss detections, respectively. 
The expected error time is defined 
in terms of the quantities in~\eqref{eq:Tspecificrate}, by
$T_{\rm err} =  \frac 1 2 \big [ T_0(-a;\gamma) + T_1(b;\gamma) \big ]$, and the error index quantifying the learning ability is its inverse, which we call the rate:
\begin{align}
r=\frac1{T_{\rm err}}  =  \frac 2 {T_0(-a;\gamma) + T_1(b;\gamma)}.
\label{eq:defTerr}
\end{align}

The second performance index $\Delta$ quantifies the adaptation ability and is again defined in terms of $T_h(z_0;z_1)$ shown in~\eqref{eq:Tgeneric}. Specifically, we consider:
\begin{align} \label{eq:Tspecificdelay}
T_1(-a;b) \qquad \textnormal{and} \qquad T_0(b;-a).
\end{align}
For the decision statistic $\bz(n)$ initialized at $\bz(0)=-a$, $T_1(-a;b)$ represents the expected time needed to hit for the first time the barrier $b$, under a steady-state state of nature~$\cH_1$. Likewise, $T_{0}(b;-a)$ is the expected time for the decision statistic $\bz(n)$, taking value~$b$ at epoch $0$, to hit for the first time the barrier~$-a$, with fixed state of nature $\cH_0$. The expected delay $\Delta$ is defined as the arithmetic mean
\begin{align}
\Delta = \frac 1 2 \Big [ T_{1}(-a;b) + T_{0}(b;-a)  \Big ] .
\label{eq:defDelta}
\end{align}

In the previous discussion, the ``typical'' value of the statistic under $\cH_0$ is $-a$, and the ``typical'' value under $\cH_1$ is $b$. These choices are natural because $-a$ and $b$ are barriers. With these choices, as we shall see soon, we obtain simple closed-form expressions for the operational characteristic $(r,\Delta)$ of the decision-maker.

However, when comparing the performance of BLLR
with that of the LMS test, sensible performance indexes for the BLLR decision-maker are obtained 
by replacing in~\eqref{eq:Tspecificrate}-\eqref{eq:defDelta} the ``typical'' values of the decision statistic under the two hypotheses, by the corresponding expected values:
\begin{align} \label{eq:means}
-a \mapsto \E_0[\bz(\infty)] \qquad \textnormal{and} \qquad b \mapsto \E_1[\bz(\infty)] ,
\end{align}
wherein we define $\bz(\infty)=\lim_{i \to \infty} \bz(i)$. The distribution of $\bz(\infty)$ is investigated, e.g., in~\cite{afanaseva}.

\subsubsection{Performance Criteria for LMS Test}
\label{sec:PCLMS}

The performance indexes of the LMS test are defined in a way similar to that of BLLR, with the notable difference that, in absence of barriers,
one cannot define the typical values of the statistic under the two hypotheses as done before, and we instead rely upon expected values.
To elaborate, assuming a steady-state hypothesis $\cH_h$, let us introduce the quantity:
\begin{align} \label{eq:TLMS}
T_h^{\prime}(w_0;w_1)= \E_h \inf_{i\ge1} \{i: \bw(i) \gtreqless w_1 ; \textnormal{ with } \bw(0)=w_0\},
\end{align}
wherein the sign $\ge$ applies if $w_0 < w_1$, and $\le$ applies if $w_0 > w_1$. 
Recall that $\{\bw(n)\}_{n\ge0}$ is defined in~\eqref{eq:LMSdet} and note the superscript $^\prime$ to distinguish quantities related to the LMS test from the corresponding quantities referring to BLLR.

As error performance indexes for LMS we consider the quantities $T_0^{\prime}(-D_{01};\gamma)$ and  $T_1^{\prime}(D_{10};\gamma)$.
The rationale is obvious. For $h=0,1$, when the state of nature is $\cH_h$ and assuming that the iteration starts from $\E_h[\bw(\infty)]$, 
we compute the expected time required to cross the threshold and therefore decide for the opposite hypothesis $\cH_{1-h}$. 
Using $T_0^{\prime}(-D_{01};\gamma)$ and  $T_1^{\prime}(D_{10};\gamma)$, we define the expected error time $T^{\prime}_{\rm err}$ as the arithmetic mean of these two quantities, and the error \emph{rate} as the inverse of $T^{\prime}_{\rm err}$:
\begin{align}
r^{\prime}=\frac1 {T^{\prime}_{\rm err}} =  \frac 2 { T^{\prime}_0(-D_{01};\gamma) + T^{\prime}_1(D_{10};\gamma) }.
\label{eq:defTerrLMS}
\end{align}

Likewise, introducing  $T^{\prime}_{1}(-D_{01};D_{10})$ and 
$T^{\prime}_{0}(D_{10};-D_{01})$, we define the expected delay as
\begin{align}
\Delta^{\prime} = \frac 1 2 \Big [ T^{\prime}_{1}(-D_{01};D_{10}) + T^{\prime}_{0}(D_{10};-D_{01}) \Big ] .
\label{eq:defDeltaLMS}
\end{align}

\subsection{Average Run Length for Page's Test}
\label{sec:ARL}

The performance of the BLLR test can be computed by borrowing results from the analysis of Page's test.
To show this, it is convenient to introduce a version $\{\bz_{\rm P}(i)\}_{i\ge0}$ of the iteration~\eqref{eq:BLLRcompact} with arbitrary starting point $c$ and a single lower barrier at $0$. This is exactly the celebrated Page's test for quickest detection~\cite{Page}:
 $\bz_{\rm P}(0)=c \ge 0$, and
\begin{align} \label{eq:page}
\bz_{\rm P}(i)=\sup \{0,\bz_{\rm P}(i-1)+ \bd(i) \} , \quad i \ge 1.
\end{align}
For $\gamma_{\rm P}>0$, 
let us define the average run length (ARL):
\begin{align}
\label{eq:ARL}
L_{h}(c;\gamma_{\rm P})= \E_h \inf_{i\ge 1} \{i:  \bz_{\rm P}(i) \ge \gamma_{\rm P}  ;  \textnormal{ with }  \bz_{\rm P}(0)=c\},
\end{align}
computed under steady-state hypothesis $\cH_h$, $h=0,1$.

In Appendix~\ref{app:exact}, an exact expression for $L_{h}(c;\gamma_{\rm P})$ is derived, involving integral equations. Since not much physical insight is gained from these integral representations, we opt for relying on approximate, but simpler and closed-form, performance formulas for $L_{h}(0;\gamma_{\rm P})$. 
These formulas are derived in Appendix~\ref{app:SPRT}, 
exploiting standard Wald's approximations~\cite{wald,wald-wolfowitz-AMS}. The final result is~\cite[Eq.\ 5.2.44]{basseville-book}:
\begin{subequations} 
\label{eq:L0001}
\begin{align}
L_{0}(0;\gamma_{\rm P})& \approx \frac{e^{\gamma_{\rm P}}-\gamma_{\rm P}-1}{D_{01}}, \\
L_{1}(0;\gamma_{\rm P})&\approx \frac{\gamma_{\rm P}+e^{-\gamma_{\rm P}}-1}{D_{10}}.
\end{align}
\end{subequations} 
For large $\gamma_{\rm P}$, expressions~\eqref{eq:L0001} simplify to:
\begin{align}
&L_{0}(0; \gamma_{\rm P})\approx \frac{e^{\gamma_{\rm P}}}{D_{01}}, \quad
L_{1}(0; \gamma_{\rm P})\approx \frac{\gamma_{\rm P}}{D_{10}} .
\end{align}

For Page's test, $L_{0}(0; \gamma_{\rm P})$ represents the mean time between false alarms and $L_{1}(0; \gamma_{\rm P})$ the worst mean delay for 
detection~\cite[Eqs.\ 5.2.18, 5.2.19]{basseville-book}.
Note the role played by the KL divergences $D_{10}$ and $D_{01}$. The larger $D_{10}$ and $D_{01}$ are, the smaller
$L_{1}(0;\gamma_{\rm P})$ and $L_{0}(0;\gamma_{\rm P})$ become, respectively. The former  has a positive impact on the performance, the latter has a negative impact.
Differently from the classical hypothesis testing problem where an increase of either or both $D_{10}$ and $D_{01}$ yields enhanced performance,
in quickest detection problem enhanced performance is obtained by increasing $D_{10}$ and/or $1/D_{01}$, as seen in~\eqref{eq:L0001}.

\section{Test Performance}
\label{sec:per}

\subsection{BLLR Test}
\label{sec:BLLR-per}

It is easy to express the performance indexes introduced in Sec.~\ref{sec:PC} in terms of the ARL $L_{h}(0;\gamma_{\rm P})$ shown in~\eqref{eq:L0001}. 
Consider first the case in which the random walk $\{\bz(i)\}_{i\ge 0}$ starts at $\bz(0)=-a$. 
For the quantities on the left of~\eqref{eq:Tspecificrate} and~\eqref{eq:Tspecificdelay} we have the obvious equalities:
\begin{align}
T_0(-a;\gamma) &= L_{0}(0;\gamma+a) = \frac{e^{\gamma+a}-(\gamma+a)-1}{D_{01}}, \label{eq:1of4}\\
T_{1}(-a;b)  &= L_{1}(0;b+a) = \frac{(b+a)+e^{-(b+a)}-1}{D_{10}} \label{eq:2of4}.
\end{align}

When the random walk $\{\bz(i)\}_{i\ge 0}$ starts at $\bz(0)=b$, we consider the reversed process $\{\bz^{-}(i)\}_{i\ge 0}$ obtained by replacing in~\eqref{eq:BLLR}
the sequence of log-likelihoods $\{\bd(i)\}_{i \ge 1}$ with its sign-reversed counterpart $\{-\bd(i)\}_{i\ge 1}$. 
Then, for the quantity $T_1(b;\gamma)$ appearing on the right of~\eqref{eq:Tspecificrate}, we have 
\begin{subequations} \label{eq:3of4}
\begin{align}
&T_1(b;\gamma)=\E_1 \inf_{i \ge 1} \{i: \bz(i) \le \gamma; \textnormal{ with }  \bz(0)=b\}, \nonumber \\
& \qquad =\E_1 \inf_{i \ge 1} \{i:  \bz^{-}(i) \ge b- \gamma;   \textnormal{ with }  \bz^{-}(0)=0\} \label{eq:symm} \\
& \qquad= L_{1}^-(0;b-\gamma) \label{eq:minus}\\
&\qquad= \frac{e^{b-\gamma}-(b-\gamma)-1}{D_{10}}, \label{eq:finqui}
\end{align}
\end{subequations} 
where~\eqref{eq:symm} follows by symmetry; the minus sign ``$^-$'' appended to the ARL in~\eqref{eq:minus} refers to a ``reversed'' Page's test in which the sequence
$\{\bd(i)\}_{i \ge 1}$ appearing in \eqref{eq:page} is replaced by $\{-\bd(i)\}_{i\ge 1}$; and~\eqref{eq:finqui} follows by noting that the ARL $L_{1}^{-}(0;\gamma_{\rm P})$
is the same of the ARL for a standard Page's test evolving under $\cH_0$ with steps whose expectation is $D_{10}$.
Similar arguments lead to 
\begin{align} \label{eq:4of4}
T_{0}(b;-a) &= L_{0}^-(0;b+a)  = \frac{(b+a)+e^{-(b+a)}-1}{D_{01}}.
\end{align}

In the case that the threshold is at the midpoint between the barriers, $\gamma=\frac{b-a}{2}$, the performance figures in~\eqref{eq:1of4}-\eqref{eq:4of4} can be expressed in terms of
the \emph{range} $R\dfz(b+a)$ of the detained random walk. 
Assuming $\gamma=\frac{b-a}{2}$, recalling the definitions of expected error time and expected delay in~\eqref{eq:defTerr} and~\eqref{eq:defDelta}, 
we get
\begin{align}
T_{\rm err} &=\frac{e^{R/2}-R/2-1}{\cD_{\rm eff}} , \label{eq:T2}\\
\Delta&=\frac{R+e^{-R}-1}{\cD_{\rm eff}} , \label{eq:DELfin}
\end{align}
where we have defined the \emph{effective divergence} between the hypotheses as
\begin{align}
\cD_{\rm eff} \dfz 2\, \frac{D_{01}D_{10}}{D_{01}+D_{10}}.
\end{align}
The inverse of $T_{\rm err}$ in~\eqref{eq:T2} is the error rate:
\begin{align}
r =\frac{\cD_{\rm eff}}{e^{R/2}-R/2-1} . \label{eq:rfin}
\end{align}
For $R \gg 1$, we obtain
\begin{align}
r  \approx \cD_{\rm eff} \, e^{-R/2}, \qquad
\Delta\approx\frac{R}{\cD_{\rm eff}} ,
\end{align}
yielding a simple insightful expression for the \emph{operational characteristic} $(r,\Delta)$ of the BLLR test: 
\begin{align}
r(\Delta)= \displaystyle{\cD_{\rm eff} \, e^{- \frac{\cD_{\rm eff} \, \Delta}{2}}}.
\label{eq:framed}
\end{align}
The function $r(\Delta)$ is strictly decreasing and convex and quantifies the fundamental trade-off of the decision procedure.
It is also worth noting that,
for a fixed $\Delta$, $r$ grows with $\cD_{\rm eff}$ as long as $\cD_{\rm eff} < \Delta/2$, while it is a decreasing function of $\cD_{\rm eff}$ for $\cD_{\rm eff} > \Delta/2$.
This behavior is to be interpreted in light of the comments reported at the end of Sec.~\ref{sec:ARL}.

\subsection{LMS Test}
Simple closed-form approximations for the test performance, similar to those shown in~\eqref{eq:1of4}-\eqref{eq:4of4}, are not available in the case of the LMS 
iteration~\eqref{eq:LMSdet}. The technical difference is that $\{\bw(i)\}_{i \ge 0}$ is not a random walk and the stopped martingale approach illustrated in 
Appendix~\ref{app:SPRT} does not apply. However, the performance of LMS can be expressed by Fredholm integral equations similar to those 
in~\eqref{eq:Fredholm}. 

To show this, we follow the approach of~\cite{crowder1987} with reference to the error figure $T^{\prime}_0(-D_{01};\gamma)$ defined in Sec.~\ref{sec:PCLMS}, see~\eqref{eq:TLMS}.  
The hypothesis in force is~$\cH_0$, the iteration $\{\bw(i)\}_{i \ge 0}$ is initialized to $\bw(0)=-D_{01}$, and the event of crossing the threshold $\gamma$
is considered. 
At the first step of the iteration, two mutually exclusive events can occur. Either $\bd(1)$ causes a threshold crossing, i.e., $\bw(1)=\mu \bd(1)+(1-\mu)\bw(0) > \gamma$, an event whose probability we denote by $p$, or $\bw(1)=\mu \bd(1)+(1-\mu)\bw(0) \le \gamma$. In the latter case the iteration restarts from $\bw(1)$ and the additional expected run length is given by $\E  T^\prime_{0}(\bw(1);\gamma)$, 
where the expectation involves the distribution of $\bd(1)$ conditioned to 
$\bd(1) \le \frac{\gamma-(1-\mu)\bw(0)}{\mu}$. Note that all distributions are computed under $\cH_0$, even if not explicitly indicated.
Let $f_{\bd|\rm cnd}(\xi)$ denote such conditional distribution, which is related to its unconditional counterpart $f_{\bd}(\xi)$  by $f_{\bd|\rm cnd}(\xi)=\frac{f_{\bd}(\xi)}{1-p}$ for $\xi \le \frac{\gamma-(1-\mu)\bw(0)}{\mu}$, and $f_{\bd|\rm cnd}(\xi)=0$ otherwise.
We obtain
\begin{align} 
&T^\prime_{0}(\bw(0);\gamma) = p + (1-p) \nonumber \\
& \qquad\quad \times \int_{-\infty}^{\infty} \Big [1+ T^\prime_{0}\big(\mu \xi+(1-\mu)\bw(0) ; \gamma \big)  \Big ] f_{\bd|\rm cnd}(\xi) \, d \xi \nonumber \\
& \, =p +\int_{-\infty}^{\frac{\gamma-(1-\mu)\bw(0)}{\mu}} \hspace*{-5pt}
\Big [1+  T^\prime_{0}\big(\mu \xi+(1-\mu)\bw(0); \gamma \big) \Big ]  f_{\bd}(\xi) \, d \xi \nonumber \\
& \, =1 + \frac 1 \mu \int_{-\infty}^{\gamma} 
T^\prime_{0}(\xi;\gamma) f_{\bd}\Big(\frac{\xi- (1-\mu) \bw(0)}{\mu}\Big) \, d \xi .
\label{eq:ARLLMS}
\end{align}
The average run length $T^\prime_{0}(\bw(0);\gamma)$ needed for the iteration $\{\bw(i) \}_{i \ge0}$ with initial value $\bw(0)$ to exceed the threshold~$\gamma$ can be computed by solving numerically~\eqref{eq:ARLLMS}.  The numerical solution to~\eqref{eq:ARLLMS} used in the examples discussed in Sec.~\ref{sec:examples} is motivated by the 
arguments provided in Appendix~\ref{app:Fredholm}. 
The quantity $T^{\prime}_{1}(-D_{01};D_{10})$ can be computed similarly to $T^{\prime}_0(-D_{01};\gamma)$, while $T^{\prime}_{1}(D_{10};\gamma)$ and $T^{\prime}_{0}(D_{10};-D_{01})$ require to consider the reversed random walk process whose steps are $\{-\bd(i)\}_{i\ge1}$. The details are omitted.

\begin{figure}
\centering 
\includegraphics[width =240pt]{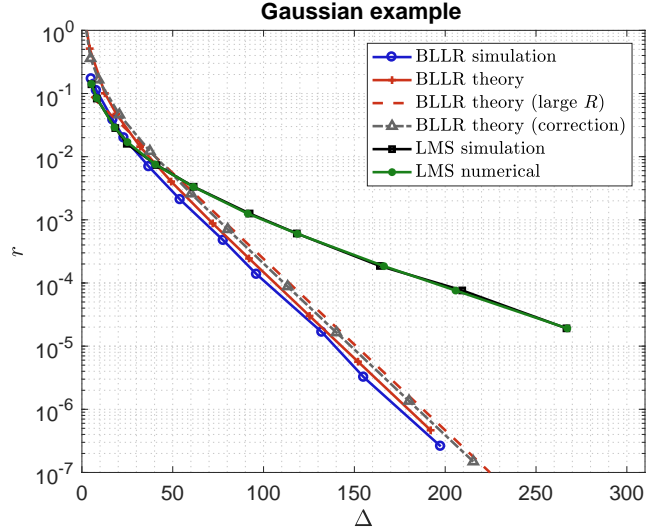}
 \caption{Gaussian example of Sec.~\ref{sec:gauss}. Operational characteristic (error rate~$r$ versus expected delay~$\Delta$) for the BLLR and LMS decision procedures. ``BLLR simulation'' shows the results of computer experiments involving $10^3$ Monte Carlo runs. ``BLLR theory'' refers to the theoretical formulas~\eqref{eq:DELfin} and~\eqref{eq:rfin}, while ``BLLR theory (large $R$)'' 
 shows the large-$R$ approximation~\eqref{eq:framed}. The curve in gray labelled as ``BLLR theory (correction)'' refers to expressions~\eqref{eq:defTerr} and~\eqref{eq:defDelta}  wherein $a$ and $b$ are replaced by the expected values as shown in~\eqref{eq:means}, for a fairer comparison with the LMS scheme. ``LMS simulation'' shows the results of computer experiments involving $10^3$ Monte Carlo runs, while the curve labelled by ``LMS numerical'' is obtained by solving numerically~\eqref{eq:ARLLMS} as detailed in Appendix~\ref{app:Fredholm}.}
      \label{fig:gauss}
\end{figure}

\section{Examples}
\label{sec:examples}

\subsection{Gaussian Shift-in-Mean}
\label{sec:gauss}
Consider the following hypotheses with IID observations $\{\bx(i)\}_{i\ge1}$: for $i=1,2,\dots$,
\begin{align}
\label{eq:testgauss}
\begin{array}{ll}
\cH_1: & \bx(i) \sim f_1(x)=\frac{1}{\sigma\sqrt{2 \pi}} e^{- \frac{(x-m)^2}{2 \sigma^2}} , \\
\cH_0: & \bx(i) \sim f_0(x)=\frac{1}{\sigma\sqrt{2 \pi}} e^{- \frac{x^2}{2 \sigma^2}}.
\end{array}
\end{align}
It is easily seen that the log-likelihood is 
\begin{equation}
\bd(i)=\log \frac{f_1(\bx(i))}{f_0(\bx(i))} = \frac{m \, \bx(i)}{\sigma^2}  - D ,
\end{equation}
where $D=D_{10}=D_{01}=\frac{m^2}{2 \sigma^2}$. The PDF of $\bd$ is $\cN(-D,2 D)$ under $\cH_0$ and  $\cN(D,2 D)$ under $\cH_1$, where $\cN(m,\sigma^2)$ denotes a Gaussian distribution with mean $m$ and variance $\sigma^2$. In the present experiment we assume $\sigma^2=1$, $m=1/2$ and $\gamma=0$, which is also the midpoint threshold because $D_{10}=D_{01}$.
For the LMS test, different values of the step-size in the range from from $8.5 \, 10^{-3}$ to $0.3$ are considered. In the case of the BLLR test,
with little loss of generality, we set the barriers as follows:
\begin{align}
a=b=\mu^{-1} D.
\end{align}
In this way, we use a single parameter $\mu$ for both LMS and BLLR decision algorithms, with the meaning that smaller values of $\mu$ imply slower adaptation.
Special attention is devoted to the slow-adaptation regime $\mu \ll1$.

Figure~\ref{fig:gauss} shows the results of computer simulations for both decision schemes. For BLLR we also show the theoretical performance shown in~\eqref{eq:DELfin} and~\eqref{eq:rfin} (theory), as well as the large-$R$ expression given in~\eqref{eq:framed} [theory (large $R$)]. For the LMS decision scheme we also show the performance obtained by solving numerically~\eqref{eq:ARLLMS} as detailed in Appendix~\ref{app:Fredholm} (LMS numerical). 
The figure confirms the accuracy of the theoretical formulas for performance prediction.

The superiority of the BLLR decision algorithm is evident, at least in the regime of small adaptation (large values of~$\Delta$). 
However, recall from the discussion in Sec.\ \ref{sec:PC} that a fair comparison between the two decision schemes requires to modify the performance indexes of the BLLR
as indicated in~\eqref{eq:means}. The expectations shown in~\eqref{eq:means} have been computed numerically and the resulting operational characteristic is shown by the gray curve in Fig.~\ref{fig:gauss}, labelled as ``theory (correction)''. 
The substantial superiority of BLLR in the low-adaptation regime is confirmed: for large $\Delta$, we see that the rate $r$ scales exponentially with the delay $\Delta$ for both the decision schemes, but the exponent is substantially larger for the BLLR decision algorithm.

\begin{figure}
\centering 
\includegraphics[width =240pt]{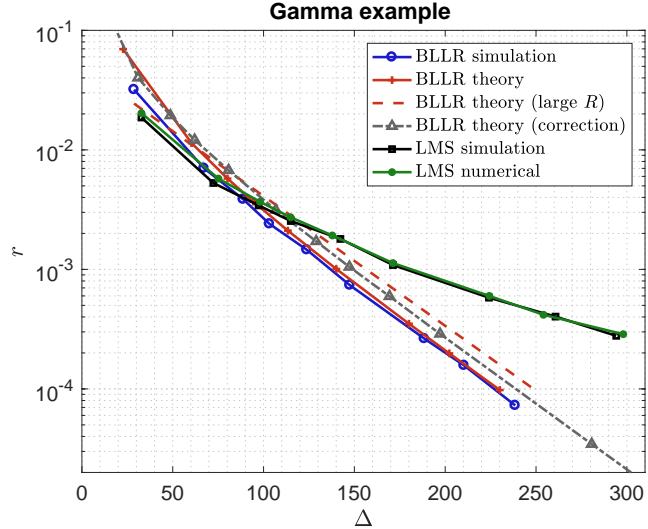}
 \caption{Example with Gamma distributions, see Sec.~\ref{sec:gauss}. The operational characteristic~$r$ versus~$\Delta$ for the BLLR and LMS decision procedures are shown.
 See the caption to Fig.~\ref{fig:gauss} for details.}
      \label{fig:gamma}
\end{figure}

\subsection{Example with Gamma Distributions}
\label{sec:gamma}
Recall the definition of the Gamma function: $\Gamma(\alpha)=\int_0^\infty \xi^{\alpha-1} e^{-\xi} \, d\xi$, with $\alpha>0$. Let $\kappa,\theta>0$. With slight abuse of notation we use the symbol $\bx \sim\Gamma(\kappa,\theta)$ to signify that $\bx$ is a Gamma-distributed random variable whose PDF is
\begin{align}
f_{\Gamma}(x)= \frac{1}{\Gamma(\kappa) \theta^{\kappa}} x^{\kappa-1}e^{-x/\theta}, \qquad x>0.
\end{align}
For $\bx \sim\Gamma(\kappa,\theta)$, it follows by straightforward algebra that $\by = \log \bx \sim {\cal L} \Gamma(\kappa,\theta)$, which is called log-Gamma distribution, having the following PDF:
\begin{align}
f_{{\cal L} \Gamma}(y)= \frac{1}{\Gamma(\kappa) \theta^{\kappa}} e^{\kappa y} e^{-\frac{e^y}{\theta}}, \qquad y>0.
\end{align}
For $y \sim {\cal L} \Gamma(\kappa,\theta)$, we have $\E y = \log \theta + \psi(\kappa)$, where $\psi(\kappa)=\frac d {dx} \log \Gamma(x)$ is known as digamma function.
We now consider the following hypotheses with IID observations $\{\bx(i)\}_{i\ge1}$ and $\rho>0$:
\begin{align}
\label{eq:testggamma}
\begin{array}{ll}
\cH_1: & \bx(i) \sim \Gamma(\kappa+\rho,\theta),\\
\cH_0: & \bx(i) \sim \Gamma(\kappa,\theta).
\end{array}
\end{align}
Simple algebra shows that the corresponding log-likelihood is $\bd(i)=\rho \log \frac{\bx(i)}{\theta} - \log \frac{\Gamma(\kappa+\rho)}{\Gamma(\kappa)}$, and therefore, with obvious notation:
\begin{align}
\label{eq:testgamma2}
\begin{array}{ll}
\cH_1: & \bd(i) \sim \rho \, {\cal L}\Gamma(\kappa+\rho,1) - \log \frac{\Gamma(\kappa+\rho)}{\Gamma(\kappa)},\\
\cH_0: & \bd(i) \sim  \rho \,  {\cal L}\Gamma(\kappa,1) - \log \frac{\Gamma(\kappa+\rho)}{\Gamma(\kappa)},
\end{array}
\end{align}
yielding 
\begin{subequations}
\begin{align}
D_{10}&=\rho \, \psi(\kappa+\rho)-\log \frac{\Gamma(\kappa+\rho)}{\Gamma(\kappa)}, \\
D_{01}&=-\rho \, \psi(\kappa)+\log \frac{\Gamma(\kappa+\rho)}{\Gamma(\kappa)}. 
\end{align}
\end{subequations}
In this experiment the barriers for the BLLR decision scheme are
\begin{align} \label{eq:bargamma}
a=\mu^{-1} D_{01}, \qquad b=\mu^{-1} D_{10}, 
\end{align}
where $\mu$ is the step-size of the LMS algorithm, and the threshold is $\gamma=\frac{b-a}{2}$.
We assume $\theta=1$, $\kappa=10$ and $\rho=1$.

The results of computer experiments for the BLLR and the LMS decision algorithms are shown in Fig.~\ref{fig:gamma}. 
The comments are similar to those of the Gaussian example. In a nutshell: the formulas for performance prediction are very accurate and the BLLR algorithm outperforms LMS, at least in the small-adaptation regime of large delays.

\begin{figure}
\centering 
\includegraphics[width =240pt]{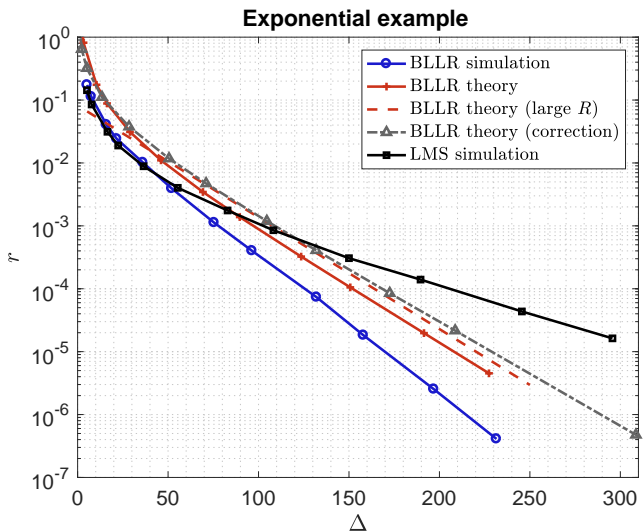}
\caption{Example with exponential distributions, see Sec.~\ref{sec:expo}. The operational characteristic~$r$ versus~$\Delta$ for the BLLR and LMS decision procedures are shown.
 See the caption to Fig.~\ref{fig:gauss} for details.}
      \label{fig:expo}
\end{figure}

\subsection{Exponentially Distributed Observations}
\label{sec:expo}
Our last example involves exponentially distributed observations: $\bx \sim {\cal E}(\eta)$ with PDF $f_{{\cal E}}(x)=\frac{e^{-x/\eta}}{\eta}$, for $x >0$ and $\eta>0$.
The two hypotheses are
\begin{align}
\label{eq:testexpo}
\begin{array}{ll}
\cH_1: & \bx(i) \sim {\cal E}(\eta_1),\\
\cH_0: & \bx(i) \sim {\cal E}(\eta_0),
\end{array}
\end{align}
with $\eta_1>\eta_0 >0$.
The corresponding log-likelihood is $\bd(i)=(\eta_0^{-1}-\eta_1^{-1}) \bx(i) -\log \frac{\eta_1}{\eta_0}$. Defining $\eta_{\rm e}=\frac{\eta_1}{\eta_0}>1$, 
for the PDFs of the likelihood $\bd$ one gets
\begin{align}
\label{eq:testexpo2}
\begin{array}{lll}
\cH_1: & f_\bd(z) = \frac{1}{\eta_{\rm e}-1} e^{-\frac{z+\log \eta_{\rm e}}{\eta_{\rm e}-1}}, & z>-\log \eta_{\rm e},\\
\cH_0: & f_\bd(z) = \frac{\eta_{\rm e}}{\eta_{\rm e}-1} e^{-\eta_{\rm e} \frac{z+\log \eta_{\rm e}}{\eta_{\rm e}-1}}, & z>-\log \eta_{\rm e} ,
\end{array}
\end{align}
and $D_{10}=\eta_{\rm e}-1-\log \eta_{\rm e}$ and $D_{01}=\eta_{\rm e}^{-1}-1+\log \eta_{\rm e}$.
In this experiment we assume $\eta_0=1$ and $\eta_1=1.5$. As for the Gamma example, the barriers for the BLLR decision scheme are
$a=\mu^{-1} D_{01}$, and $b=\mu^{-1} D_{10}$, where $\mu$ is the step-size of the LMS algorithm, and we use the mid-point threshold $\gamma=\frac{b-a}{2}$.

The results of computer simulations compared to the theoretical formulas are shown in Fig.~\ref{fig:expo}. The comments are similar to the previous case, but in the exponential case
the theoretical formulas are less accurate. The slope of the operational curve $r(\Delta)$ seems correctly predicted by the analytical formulas, 
but a multiplicative correction appears to be necessary. This is a manifestation of the poor accuracy of Wald's approximations of neglecting the excess over the boundaries, which have been exploited to derive the theoretical formulas. Improvements in this regard are possible, e.g., via nonlinear renewal theory~\cite[Sec.\ 2.6]{tartakovsky-book},~\cite{siegmund}, but not pursued here.
In addition, in the exponential case, the theoretical operational characteristic of the LMS decision scheme is not reported because of instability of the numerical procedure detailed in Appendix~\ref{app:Fredholm} to solve~\eqref{eq:ARLLMS}. 

\section{An Application Related to COVID-19 Pandemic}
\label{sec:covid}

During the course of a pandemic, one of the most challenging tasks for authorities is to decide when to impose or relax restrictive measures with huge societal and economic costs, such as: closure of schools, universities, shops, factories, limitation of social activities, strict lockdown. In this respect, learning and adaptation algorithms can be useful to support informed and rational decision making. In this section, we discuss a variation of the BLLR test, which is particularly relevant in connection to the analysis of COVID-19 pandemic time-series.

Let us start by considering the classical SIR model of pandemic evolution introduced by Kermack and McKendrick in 1927~\cite{SIR-1927}. Let\footnote{We adopt a standard notation for the SIR model. Thus, in this section~$r(t)$ denotes the fraction of recovered individuals, not to be confused with the rate~$r$ introduced in Sec.~\ref{sec:PC}. The fraction of infected individuals is denoted by~$i(t)$ and should not be confused with the time index~$i$. Finally, the recovering rate should not be confused with the threshold~$\gamma$ previously introduced. The differences should be clear from the context.} $s(t)$, $i(t)$ and $r(t)$ denote the fractions of susceptible, infected, and recovered (or dead) individuals, respectively. Let $\beta$ be the infection rate (infected individuals per unit time), and $\gamma$ the recovering rate.
The celebrated SIR equations are~\cite{allen-2017-primer}:
\begin{align} \label{eq:SIR}
\begin{cases}
\frac{ds(t)}{dt} &= - \beta \, s(t) i(t), \\
\frac{di(t)}{dt} &= \beta \, s(t) i(t) - \gamma \, i(t), \\
\frac{dr(t)}{dt} &=  \gamma \, i(t),
\end{cases} 
\end{align}
with the initial conditions $r(0)=0$, $s(0)=1-i(0)$. We assume $0< i(0) \ll 1$, where $i(0)$ represents the small fraction of the total population from which the infection originates.
Let us focus on the situation in which the pandemic is mostly under control, because of restrictions imposed by the authorities such as social distancing, but at the same time is not eradicated. Then, it is reasonable to assume that the fraction of susceptible individuals is maintained almost constant $s(t) \approx s_\ast$, implying that the second equation in~\eqref{eq:SIR} reduces to 
\begin{align} \label{eq:expoi}
\frac{di(t)}{dt} = (\beta s_\ast-\gamma) \, i(t) .
\end{align}
Since data about the infections are typically collected on a daily basis, consider a discrete-time version of \eqref{eq:expoi}
with unit-step discretization (we loosely use the same notation $i(\cdot)$ for the time-discrete version): 
\begin{align}
&\Delta i(k) \dfz i(k)- i(k-1) =(\beta s_\ast-\gamma) \, i(k-1) \label{eq:idiscr}\\
&\quad \Rightarrow \quad i(n)= i(0) (1+\beta s_\ast-\gamma)^n,
\end{align}
for some $i_0>0$. From \eqref{eq:idiscr} we see that the ratio $i(k)/i(k-1)$ is constant.
It is evident that real-world data are ``noisy'' versions of the previous deterministic equations. Accordingly, we model the ratio $\bi(k)/\bi(k-1)\dfz \bx(k)$
as a random variable. Precisely, we assume:
\begin{align} \label{eq:model}
\bi(n)= \bi(0) \prod_{k=1}^n \bx(k), \qquad n \ge 1, 
\end{align}
where $\{\bx(k)\}_{k \ge 1}$ is a sequence of \emph{independent} random variables. 
In light of~\eqref{eq:model}, $\bx(k)$ is referred to as \emph{growth rate}.

\begin{figure}
\centering 
\includegraphics[width =240pt]{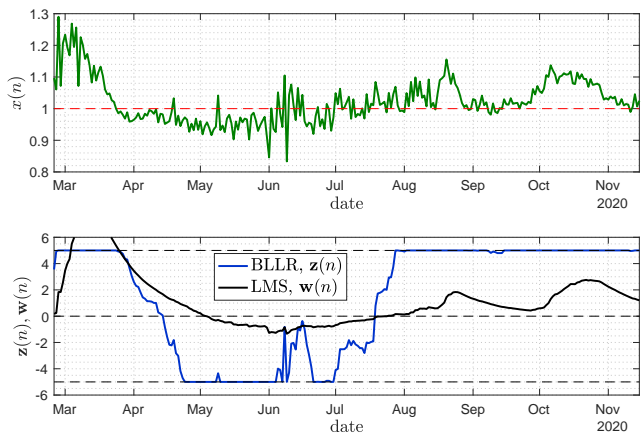}
\caption{\emph{Top:} The sequence of growth rates $\{x(n)\}$ of new positive individuals in Italy, during the COVID-19 pandemic. \emph{Bottom:} Decision statistics obtained by running the BLLR and LMS procedures on the sequence $\{x(n)\}$. Decision is for $\cH_1$ when the statistic is positive and for $\cH_0$ otherwise. We use $\sigma=0.036$, $a=b=5$ and $\mu=0.05$.}
      \label{fig:covid}
\end{figure}

By exploiting the publicly available data of the COVID-19 illness spread in Italy (freely downloadable at https://github.com/pcm-dpc/COVID-19/), we obtain\footnote{Precisely, the procedure is as follows. The sequence of the new positives per day $\{p(k)\}_{k\ge 1}$ is downloaded and smoothed by a moving mean filter of length $7$ days, to clean gross errors from data. Then, the growth rate  is computed as $x(k)=p(k+1)/p(k)$, which is the same as $x(k)=i(k)/i(k-1)$, see~\cite{MAST-NC,MAST-SPL}.} the sequence of growth rates $\{\bx(k)\}_{k\ge 1}$ and verify that the $\bx(k)$'s are well-represented by Gaussian random variables;
details can be found in~\cite{MAST-NC,MAST-SPL}.
The expected value $\E \bx(k)$ is time-varying and unknown, and characterizes the specific phase of the pandemic: when the pandemic is under control --- a situation here referred to as hypothesis $\cH_0$ --- we have $m_{0}(k) \dfz \E_0 \bx(k) \le 1$. Conversely, under the alternative  hypothesis~$\cH_1$, $m_{1}(k)\dfz \E_1 \bx(k) > 1$ and the contagion grows exponentially fast.

Lacking knowledge of the sequences of the mean values $\{m_{0}(k)\}_{k \ge 1}$ and $\{m_{1}(k)\}_{k \ge 1}$, one cannot compute the log-likelihood in~\eqref{eq:LL} and the related BLLR statistic in~\eqref{eq:BLLR}. We then resort to a GLRT approach~\cite{poorbook}, which amounts to replacing the unknown parameters appearing in the log-likelihood with their ML estimates. 
The approach is similar to that pursued in~\cite{Sayed-CooperativeSensing} in the context of SPRT problems, but the estimates are structurally different.
In our case the number of unknown parameters  grows linearly in time and the estimates are constrained to $\widehat m_{0}(k) \le 1$ and $\widehat m_{1}(k)>1$, for all $k$. It is simple to see that

\vspace*{-10pt}
{\small \begin{align}
    &\widehat m_{0}(k)= \arg\max_{m \le 1}{\frac{1}{\sqrt{2 \pi \sigma^2}} e^{-\frac{(\bx(k)-m)^2}{2 \sigma^2}}=\min(\bx(k),1)}, 
    \label{eq:ML0}\\
    &\widehat m_{1}(k)= \arg\max_{m > 1}{\frac{1}{\sqrt{2 \pi \sigma^2}} e^{-\frac{(\bx(k)-m)^2}{2 \sigma^2}}=\max(\bx(k),1)}. 
    \label{eq:ML1}
\end{align}}%
Computing the log-likelihood yields:
{\small \begin{align}
\log \frac{f_1(\bx(k))}{f_0(\bx(k))} 
=\frac 1 {2 \sigma^2}  [(\bx(k)-m_0(k))^2-(\bx(k)-m_1(k))^2].
\label{eq:LLgaus}
\end{align}}%
Substituting the ML estimates~\eqref{eq:ML0} and~\eqref{eq:ML1} in place of $m_{0}(k)$ and $m_{1}(k)$ in~\eqref{eq:LLgaus}, one obtains
\begin{align}
    \bd(k)=\frac 1 {2 \sigma^2}
    (\bx(k)-1)^2 \sign(\bx(k)-1).
    \label{eq:newd}
\end{align}

Using $\bd(k)$ shown in~\eqref{eq:newd}, in place of~\eqref{eq:LL}, we get the following ``GLRT'' version of BLLR: 
$\bz(0)=0$ and
\begin{align}
\bz(k)=\inf \big \{ b, \sup \{-a,\bz(k-1)+ \bd(k) \} \big \}, \quad k \ge 1.
\label{eq:BLLRcompactnew}
\end{align}
Clearly, the definition of $\bd(k)$ in~\eqref{eq:newd} can also be used in the LMS iteration~\eqref{eq:LMSdet} to obtain a ``GLRT'' version of LMS. These ``GLRT'' versions are exactly as described in Algorithms~\ref{alg:BLLR} and~\ref{alg:LMS}, but take a different sequence$\{\bd(k)\}$ in input.


\begin{figure}
\centering 
\includegraphics[width =200pt]{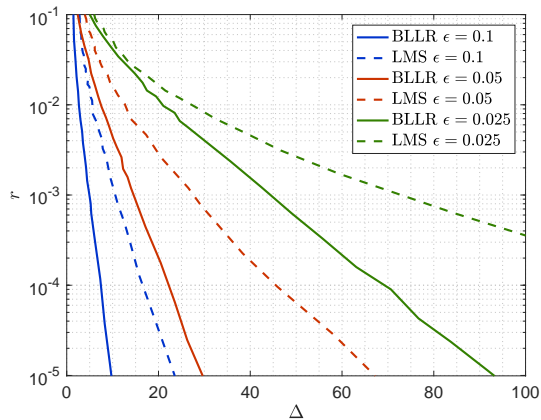}
\caption{Operational characteristics of the ``GLRT'' versions of BLLR and LMS,
relevant to COVID-19 pandemic control.
The mean values of the growth rate are drawn uniformly from $(1-\epsilon,1)$ under $\cH_0$ and
uniformly from $(1,1+\epsilon)$ under $\cH_1$. We use $10^3$ Monte Carlo runs for each value of $a=b$ (BLLR) and $\mu$ (LMS), mid-point (zero) threshold and $\sigma=0.036$.}
      \label{fig:covidoc}
\end{figure}

We now demonstrate the GLRT versions of BLLR and LMS on COVID-19 time-series recorded in Italy. Top panel of Fig.~\ref{fig:covid} shows the growth rate sequence $\{x(n)\}_{n\ge 1}$, where~$n$ denotes the index of the day corresponding to the date on the abscissa, from February 25 to November 15, 2020. Bottom panel shows the BLLR and LMS decision statistics computed from $\{x(n)\}_{n\ge 1}$. The standard deviation appearing in~\eqref{eq:newd}, estimated from the data, is $\sigma = 0.036$. For BLLR, we use $a=b=5$; for LMS, $\mu=0.05$. Assuming zero threshold for both decision statistics, in the case of BLLR we see that passage $\cH_1 \mapsto \cH_0$ is declared on April 15, followed by passage  $\cH_0 \mapsto \cH_1$ on July 18, while, for LMS, the former is declared on May 4, and the latter on July 24. We see that BLLR reacts more promptly to the change of state of nature.

To investigate the performance of BLLR and LMS on COVID-19 pandemic data, we compute~$\Delta$ and~$r$ by computer experiments, assuming that the mean values of the data are IID random variables uniform in $(1-\epsilon,1)$ under $\cH_0$, and uniform in $(1,1+\epsilon)$ under $\cH_1$, for some $0< \epsilon < 1$. Given the mean value, Gaussian data are generated with standard deviation  $\sigma=0.036$. In the case of BLLR, we explore various values of $a=b$. In the case of LMS, we compute numerically $\E_1 \bd =-\E _0 \bd$, see~\eqref{eq:twodiv}, and explore various values of $\mu$. In both cases, $10^3$ Monte Carlo runs are used and the mid-point threshold $0$ is selected. The resulting operational characteristics are depicted in Fig.~\ref{fig:covidoc}. We see that BLLR
outperforms LMS, consistently with the results of Sec.~\ref{sec:examples}.

\section{Conclusion}
\label{sec:conclusion}

Learning and adaptation algorithms have been originally designed for solving estimation tasks. Then, they have been exploited in decision contexts to compute an efficient online estimation of the optimal decision statistic. Thus, it seemed natural to build the decision system on the same LMS component used for estimation purposes. In this paper, we focus on learning and adaptation schemes for solving decision problems. We propose an alternative to LMS, called BLLR, as core element of the decision system. Performance analysis reveals that BLLR can outperform LMS, in the slow-adaptation regime. 
Our study is limited to a single decision maker and paves the way to further investigations aimed at designing the diffusion step for a network of interconnected decision makers, much in the same way as the ATC diffusion rule has been advocated to be used in combination with LMS.

An application to COVID-19 pandemic data has been discussed, using a ``GLRT'' version of BLLR. Elaborating on the time-series of daily new positive individuals from Italy in the period from February 25 to November 15, 2020, we show that the proposed approach effectively tracks changes of pandemic phases, providing a rigorous tool to quickly detect the passage from a controlled regime~$\cH_0$ in which the number of new positives tends to decrease or be stable, to a critical regime~$\cH_1$ of pandemic explosion, and vice versa.
In line with the considerations made in~\cite{MAST-NC,MAST-SPL}, aside from the growth rate of new positives, the BLLR algorithm can be executed on other pandemic time-series (e.g., hospitalizations, ratio positive/tested individuals, etc.), 
paving the way to more comprehensive analyses of COVID-19 data.

\begin{appendices}
\numberwithin{equation}{section}

\section{Exact Solutions for the ARL}
\label{app:exact}

Page's test can be regarded as a set of parallel open-ended Wald's SPRTs (see~\cite{wald,wald-wolfowitz-AMS}). The following analysis is based on this similitude, see~\cite[Sec. 5.2.2.1]{basseville-book}. Setting $c=0$ in~\eqref{eq:ARL}, $L_{h}(0;\gamma_{\rm P})$ can be written as the sum of two contributions. The first is the average sample number (ASN) of the SPRT for $i\ge 1$ given that the SPRT hits back the level $0$ before crossing $\gamma_{\rm P}$, multiplied the expected number of such ``back to $0$'' cycles (a random quantity which is easily seen to follow a geometric distribution with range $\{0,1,2\dots\}$, because any time the process hits $0$ it probabilistically restarts). The second contribution is the ASN of the SPRT given that it crosses the threshold $\gamma_P$ before hitting back the $0$ level. 
Denoting by $p_0$ the probability of hitting zero before crossing $\gamma_P$ (starting from 0), we see that the expected number of back-to-$0$ cycles is $p_0/(1-p_0)$. 
Then, denoting by $\E_h \bn_0$ the ASN of the SPRT we have:
\begin{align}
L_{h}(0;\gamma_{\rm P})&= \E_h[\bn_0|\textnormal{back to }0]\frac{p_0}{1-p_0} + \E_h[\bn_0|\textnormal{cross } \gamma_{\rm P}] \nonumber \\
&= \frac{\E_h[\bn_0|\textnormal{back to }0]p_0+\E_h[\bn_0|\textnormal{cross } \gamma_{\rm P}] (1-p_0)}{1-p_0} \nonumber \\
&=\frac{\E_h \bn_0 }{1-p_0}.
\label{ARL0}
\end{align}

Using $L_{h}(0;\gamma_{\rm P})$, a similar equation for the case of $c>0$ can also be derived. For the SPRT with lower threshold at 0, let $p_c$ be the probability of hitting zero before crossing $\gamma_{\rm P}$, when the random walk starts from $\bz_{\rm P}(0)=c$, and let $\E_h \bn_c$ be the corresponding ASN. Starting from $c$, two mutually exclusive situations may occur: either $ \bz_{\rm P}(i)$ hits zero before crossing $\gamma_{\rm P}$, which happens with probability $p_c$, or its complement occurs and $\bz_{\rm P}(i)$ crosses $\gamma_{\rm P}$ without hitting zero. In the former case the average run length equals the time needed to hit 0 plus $L_{h}(0;\gamma_{\rm P})$. In formula:
\begin{align}
L_{h}(c;\gamma_{\rm P})&=  p_c \Big [ \E_h[\bn_c|\textnormal{back to }0]+L_{h}(0;\gamma_{\rm P})  \Big ] \nonumber \\
& + (1-p_c) \, \E_h[\bn_c|\textnormal{cross } \gamma_{\rm P}]  \nonumber \\
&= p_c L_{h}(0; \gamma_{\rm P}) + \E_h \bn_c . 
\label{ARLc}
\end{align}

It can be shown that $p_c$ and $\E_h \bn_c$ are  solutions to the following Fredholm integral equations of the second kind~\cite[p.\ 168]{basseville-book}:
for $0 \le c \le \gamma_{\rm P}$:
\begin{subequations} 
\label{eq:Fredholm}
\begin{align}
p_c&= \int_{-\infty}^{-c} f_\bd(\xi) \, d\xi + \int_0^{\gamma_{\rm P}} p_\xi f_\bd(\xi-c) \, d\xi \\
\E_h \bn_c &= 1+ \int_0^{\gamma_{\rm P}} \E_h \bn_\xi \; f_\bd(\xi-c) \, d\xi ,
\end{align}
\end{subequations} 
where $f_\bd(\xi)$ is the PDF of the log-likelihood $\bd(i)$ shown in~\eqref{eq:LL}, under hypothesis $\cH_h$ (dependence not made explicit for notational simplicity). Among others, iterative methods have been proposed for solving these equations, see e.g.,~\cite[Eq.\ 5.2.28]{basseville-book}. 
After computing $p_c$ and $\E_h \bn_c$, we obtain $L_{h}(c;\gamma_{\rm P})$ from~\eqref{ARL0} and~\eqref{ARLc}, and the ARL is found.

\section{Classical Results on SPRTs using Martingales}
\label{app:SPRT}
Suppose the state of nature is $\cH_0$. Consider the iteration in~\eqref{eq:walkz}: $\bz_{\rm S}(0)=0$ and 
\begin{align}
\bz_{\rm S}(i)=\bz_{\rm S}(i-1)+ \bd(i), \quad i \ge 1,
\end{align}
where $\bd(i)$ is the log-likelihood of the $i$-th observation defined in~\eqref{eq:LL}, and we have appended an ``${\rm S}$'' to indicate that we now
consider a SPRT. Recall that observations $\{\bx(i)\}_{i\ge1}$ are IID. Clearly, $\bz_{\rm S}(n)=\sum_{i=1}^n \bd(i)$. The random process $e^{\bz_{\rm S}(n)}$ is called a martingale with respect to the sequence $\{\bd(i)\}_{i \ge 1}$ because:
\begin{align}
\E_0 [e^{\bz_{\rm S}(n)} | \bd(n-1),\dots, \bd(1)]&= e^{\bz_{\rm S}(n-1)} \E_0 e^{\bd(n)} \nonumber \\
&= e^{\bz_{\rm S}(n-1)} ,
\end{align}
and because the regularity condition $\E_0 |e^{\bz_{\rm S}(n)}|<\infty$, $\forall n$, is obviously met.
For some pair of thresholds $-\gamma_\ell < 0 < \gamma_u$, let
\begin{align}
\bn = \inf_{i \ge 1} \{ i: \bz_{\rm S}(i) \ge \gamma_u \textnormal{ or } \bz_{\rm S}(i) \le -\gamma_\ell\}
\end{align}
be a random time for the process $e^{\bz_{\rm S}(n)}$. It can be shown that $\P_0 (\bn < \infty)=1$, e.g., by considering that
the sequence of descending ladder heights of $e^{\bz_{\rm S}(n)}$ is non terminating, see~\cite{FellerBookV2}. Because $\bn$ is finite with probability one, $\bn$ is a \emph{stopping time}, and the expected value of the martingale at the stopping time equals the expected value at the initial time~\cite[Th. 6.2.2]{ross-book}:
\begin{align}
\E_0[e^{\bz_{\rm S}(\bn)}]=\E_0[e^{\bz_{\rm S}(1)}]=\E_0\bigg[\frac{f_1(\bx)}{f_0(\bx)}\bigg]=1.
\label{eq:=1}
\end{align}
Using the so-called Wald's approximations of neglecting the excesses over the boundaries, namely,
\begin{subequations}
\label{eq:neob}
\begin{align}
&\E_0[e^{\bz_{\rm S}(\bn)} | \bz_{\rm S}(\bn) \ge \gamma_u] \approx e^{\gamma_u}, \\
&\E_0[e^{\bz_{\rm S}(\bn)} | \bz_{\rm S}(\bn) \le - \gamma_l] \approx e^{-\gamma_l},
\end{align}
\end{subequations}
from~\eqref{eq:=1} we arrive at
\begin{align}
1&=\E_0[e^{\bz_{\rm S}(\bn)} | \bz_{\rm S}(\bn) \ge \gamma_u] \Big (1-\P_0(\bz_{\rm S}(\bn) \le -\gamma_\ell) \Big) \nonumber \\
&\quad + \E_0[e^{\bz_{\rm S}(\bn)} | \bz_{\rm S}(\bn) \le -\gamma_\ell] \P_0(\bz_{\rm S}(\bn) \le -\gamma_\ell) \nonumber \\
&\approx e^{\gamma_u}\Big (1-\P_0(\bz_{\rm S}(\bn) \le -\gamma_\ell) \Big)+e^{-\gamma_\ell} \P_0(\bz_{\rm S}(\bn) \le -\gamma_\ell),
\end{align}
or
\begin{align}
\P_0(\bz_{\rm S}(\bn) \le -\gamma_\ell) \approx \frac{e^{\gamma_u}-1}{e^{\gamma_u}- e^{-\gamma_\ell}}.
\label{eq:first}
\end{align}

Assuming $\E_0|\bd|< \infty$ and $\E_0\bn<\infty$ we have (Wald's equation)~\cite[Th. 3.3.2]{ross-book}:
\begin{align}
\E_0[\bz_{\rm S}(\bn)] = \E_0\bd \, \E_0\bn = - D_{01} \, \E_0\bn .
\end{align}
Using approximations similar to those in \eqref{eq:neob}, this yields
\begin{align}
\E_0[\bn] &\approx \frac{\gamma_u \Big ( 1-\P_0(\bz_{\rm S}(\bn) \le - \gamma_\ell) \Big ) - \gamma_\ell  \P_0(\bz_{\rm S}(\bn) \le - \gamma_\ell)}{- D_{01}} \nonumber \\
& \, = \frac{\P_0(\bz_{\rm S}(\bn) \le - \gamma_\ell) (\gamma_u + \gamma_\ell) - \gamma_u }{D_{01}}.
\label{eq:second}
\end{align}

Consider now expression~\eqref{ARL0} under $\cH_0$, namely for $h=0$:
\begin{align}
L_{0}(0;\gamma_{\rm P})  =\frac{\E_0 \bn_0}{1-p_0}.
\label{ARL00}
\end{align}
By setting $\gamma_u =\gamma_{\rm P}$ and taking the limit $\gamma_\ell \to 0$, $\E_0 \bn$ becomes $\E_0 \bn_0$, $\P_0(\bz_{\rm S}(\bn) \le -\gamma_\ell)$
reduces to $p_0$, and the quantity in~\eqref{ARL00} can be computed by exploiting relationships~\eqref{eq:first} and~\eqref{eq:second}:
\begin{align}
L_{0}(0;\gamma_{\rm P})  &= \frac 1 {D_{01}} \lim_{\gamma_\ell \to 0} \frac{\frac{e^{\gamma_{\rm P}}-1}{e^{\gamma_{\rm P}}- e^{-\gamma_\ell}} (\gamma_{\rm P} + \gamma_\ell) - \gamma_{\rm P} }{\frac{1-e^{-\gamma_\ell}}{e^{\gamma_{\rm P}}- e^{-\gamma_\ell}}} \nonumber \\
&= \frac{e^{\gamma_{\rm P}}-\gamma_{\rm P}-1}{D_{01}} .
\end{align}

Similar calculations hold under $\cH_1$, considering this time the martingale $e^{-\bz_{\rm S}(n)}$, yielding
\begin{align}
L_{1}(0;\gamma_{\rm P})  &=  \frac{\gamma_{\rm P}+e^{-\gamma_{\rm P}}-1}{D_{10}} .
\end{align}

\section{Numerical Solutions to \eqref{eq:ARLLMS}}
\label{app:Fredholm}

In many cases of interest, an approximate numerical solution to~\eqref{eq:ARLLMS} can be obtained by a simple iterative procedure,
based on heuristic arguments. 
Let us start by considering a modified version of the problem in which the iteration $\{\bw(i)\}_{i \ge 0}$ with initial point $\bw(0)=-D_{01}$ evolves up to cross one of the \emph{two} thresholds $\gamma_\ell<-D_{01}$ and $\gamma>-D_{01}$. The same arguments used to derive~\eqref{eq:ARLLMS} lead to the Fredholm integral equation
\begin{align}
T^\prime_{0}(-D_{01};\gamma) = 1 \hspace*{-2pt} + \hspace*{-2pt} \frac 1 \mu \hspace*{-2pt} \int_{-\gamma_\ell}^{\gamma} \hspace*{-5pt}
T^\prime_{0}(\xi;\gamma) f_{\bd}\Big(\frac{\xi+ (1-\mu) D_{01}}{\mu}\Big) \, d \xi ,
\label{eq:ARLLMSmod}
\end{align} 
wherein $\mu$ is the step-size and $f_{\bd}(\cdot)$ is the distribution of~$\bd$ under $\cH_0$. Let us consider an operator ${\cal P}: {\cal C}[-\gamma_\ell,\gamma]
\mapsto {\cal C}[-\gamma_\ell,\gamma]$ mapping the complete metric (Banach) space of continuous functions ${\cal C}[-\gamma_\ell,\gamma]$ defined over the closed interval $[-\gamma_\ell,\gamma]$, into itself. The operator ${\cal P}$ is defined by:
\begin{align}
{\cal P}(g(x)) = 
1 + \frac 1 \mu \int_{-\gamma_\ell}^{\gamma} 
g(\xi) f_{\bd}\Big(\frac{\xi- (1-\mu) x}{\mu}\Big) \, d \xi .
\end{align}
As seen in~\eqref{eq:ARLLMSmod}, $T^\prime_{0}(x;\gamma)$, $x \in [-\gamma_\ell,\gamma]$ is a fixed point of the operator ${\cal P}$, namely
${\cal P}(T^\prime_{0}(x;\gamma))=T^\prime_{0}(x;\gamma)$. Thus,  
the problem of solving~\eqref{eq:ARLLMSmod} reduces to the problem of finding such fixed point, provided that it exists and is unique.

Picard-Banach fixed point principle states that a mapping ${\cal P}$ of a complete metric space into itself has a unique fixed point $g^\ast$ provided that the mapping is a 
\emph{contraction}~\cite[Sec. 9.7]{Zorichbook2}. The operator ${\cal P}$ is called a contraction if 
\begin{align}
\| {\cal P}(g_1) - {\cal P}(g_2) \| \le \delta \| g_1 - g_2 \|,
\label{eq:defcontract}
\end{align}
for some $0 < \delta < 1$. In addition, for a contraction, the iteration 
\begin{align} 
g_{n}={\cal P}(g_{n-1}), \qquad n \ge 1,
\label{eq:iterg}
\end{align} 
converges to the fixed point $g^\ast$ with a rate of convergence 
\begin{align}
\| g_n - g^\ast\| \le \frac{\delta^n}{1-\delta} \| g_1 - g_0\|,
\end{align}
where $g_0 \in {\cal C}[-\gamma_\ell,\gamma]$ is some initial guess~\cite[Eq. (9.21)]{Zorichbook2}. 

Thus, \eqref{eq:iterg} provides a simple numerical recipe to solve~\eqref{eq:ARLLMSmod}.
To check that ${\cal P}$ is a contraction, using the supremum norm, note that
\begin{subequations}
\begin{align}
& \| {\cal P}(g_1) - {\cal P}(g_2) \| = \sup_{x\in[-\gamma_\ell,\gamma]} \Big| {\cal P}(g_1(x)) - {\cal P}(g_2(x)) \Big| \nonumber \\
& = \frac 1 \mu \sup_{x\in[-\gamma_\ell,\gamma]} \Bigg |  \int_{-\gamma_\ell}^{\gamma} 
\Big [g_1(\xi) -g_2(\xi) \Big ] f_{\bd}\Big(\frac{\xi- (1-\mu) x}{\mu}\Big) \, d \xi \Bigg | \nonumber \\
& \le \frac 1 \mu \sup_{x\in[-\gamma_\ell,\gamma]}  
\int_{-\gamma_\ell}^{\gamma} 
\Big | g_1(\xi) -g_2(\xi) \Big | f_{\bd}\Big(\frac{\xi- (1-\mu) x}{\mu}\Big) \, d \xi \nonumber \\
& =  \Big | g_1(x^\ast) -g_2(x^\ast) \Big | \sup_{x\in[-\gamma_\ell,\gamma]} 
\int_{-\gamma_\ell}^{\gamma}  \frac 1 \mu f_{\bd}\Big(\frac{\xi- (1-\mu) x}{\mu}\Big) \, d \xi \label{eq:mvt} \\
& \le  \Big | g_1(x^\ast) -g_2(x^\ast) \Big | \le \sup_{x\in[-\gamma_\ell,\gamma]}  \Big | g_1(x) -g_2(x) \Big | \label{eq:pl1} \\
&  = \| g_1(x) -g_2(x) \| ,
\label{eq:contraction}
\end{align}
\end{subequations}
where in~\eqref{eq:mvt} $x^\ast \in [-\gamma_\ell,\gamma]$ and the equality follows by the mean value theorem for integrals~\cite[Th. 5, p.352]{Zorichbook}, while
the first inequality in~\eqref{eq:pl1} follows because the integrand in~\eqref{eq:mvt} represents a valid PDF. 


The rigorous approach provided by the theory of fixed points does not apply directly to our case because in~\eqref{eq:ARLLMS} 
we have $\gamma_\ell \to \infty$ and the derivation leading to~\eqref{eq:contraction} fails for unbounded intervals, in which the mean value theorem for integrals do not apply.
Numerical approaches to the solution of Fredholm integral equations defined over unbounded intervals have been proposed, see~\cite{Avazzadeh} and the references therein. 
However, for the sake of simplicity, here we limit ourselves to apply~\eqref{eq:iterg}, checking numerically the convergence.
Summarizing, we use the following heuristic approach: start with some initial guess $g_0(x)$, $x\in (-\infty,\gamma]$, and iterate for $n\ge 1$:
\begin{align}
g_{n}(x) = 1  +  \frac 1 \mu  \int_{-\infty}^{\gamma} \hspace*{-5pt}
g_{n-1}(\xi) f_{\bd}\Big(\frac{\xi- (1-\mu) x}{\mu}\Big) \, d \xi 
\label{eq:heuristic}
\end{align}
up to convergence, if any. In the computer experiments presented in this work the initial guess $g_0(x)$ is taken as a constant function, whose value is 
equal to the value $T^\prime_{0}(-D_{01};\gamma)$ obtained by simulations. 
The numerical evaluations of the performance for the LMS decision algorithm presented in this paper are based on~\eqref{eq:heuristic} and similar iterates, using numerical integration for integrals.

\end{appendices}


\begin{thebibliography}{10}
\providecommand{\url}[1]{#1}
\csname url@samestyle\endcsname
\providecommand{\newblock}{\relax}
\providecommand{\bibinfo}[2]{#2}
\providecommand{\BIBentrySTDinterwordspacing}{\spaceskip=0pt\relax}
\providecommand{\BIBentryALTinterwordstretchfactor}{4}
\providecommand{\BIBentryALTinterwordspacing}{\spaceskip=\fontdimen2\font plus
\BIBentryALTinterwordstretchfactor\fontdimen3\font minus
  \fontdimen4\font\relax}
\providecommand{\BIBforeignlanguage}[2]{{%
\expandafter\ifx\csname l@#1\endcsname\relax
\typeout{** WARNING: IEEEtran.bst: No hyphenation pattern has been}%
\typeout{** loaded for the language `#1'. Using the pattern for}%
\typeout{** the default language instead.}%
\else
\language=\csname l@#1\endcsname
\fi
#2}}
\providecommand{\BIBdecl}{\relax}
\BIBdecl


\bibitem{chamberland}
J.-F. Chamberland and V.~V. Veeravalli, ``Decentralized detection in sensor
  networks,'' \emph{{IEEE} Trans. Signal Process.}, vol.~51, no.~2, pp.
  407--416, Feb. 2003.

\bibitem{PreddKulkarniPoor-magazine06}
J.~B. Predd, S.~R. Kulkarni, and H.~V. Poor, ``Distributed learning in wireless
  sensor networks,'' \emph{{IEEE} Signal Process. Mag.}, vol.~23, no.~4, pp.
  56--69, Jul. 2006.

\bibitem{akyildiz-survey}
J.~Akyildiz, W.~Su, Y.~Sankarasubramaniam, and E.~Cayirci, ``A survey on sensor
  networks,'' \emph{{IEEE} Commun. Mag.}, vol.~40, no.~8, pp. 102--114, Aug.
  2002.

\bibitem{Varshney:book}
P.~K. Varshney, \emph{Distributed Detection and Data Fusion}. New York, NY: Springer, 1997.

\bibitem{viswanathan97}
R.~Viswanathan and P.~K. Varshney, ``Distributed detection with multiple
  sensors: Part {I} -- fundamentals,'' \emph{Proc. {IEEE}}, vol.~85, no.~1, pp.
  54--63, Jan. 1997.

\bibitem{blum97}
R.~S. Blum, A.~Kassam, and H.~V. Poor, ``Distributed detection with multiple
  sensors: Part {II} -- advanced topics,'' \emph{Proc. {IEEE}}, vol.~85, no.~1,
  pp. 64--79, Jan. 1997.

\bibitem{tsitsiklis93}
J.~N. Tsitsiklis, ``Decentralized detection,'' in \emph{Advances in Signal
  Processing}, H.~V. Poor and J.~B. Thomas, Eds., JAI Press, 1993, pp. 297--344.

\bibitem{Tsitsiklis88}
------, ``Decentralized detection by a large number of sensors,'' \emph{Math.
  Contr., Signals, Syst.}, vol.~1, pp. 167--182, 1988.

\bibitem{SENMA}
L.~Tong, Q.~Zhao, and S.~Adireddy, ``Sensor networks with mobile agents,'' in
  \emph{Proceedings of {MILCOM} 2003}, vol.~1, Boston MA, Oct. 2003, pp.
  688--693.

\bibitem{doasplet05}
S.~Marano, V.~Matta, P.~Willett, and L.~Tong, ``{DOA} estimation via a network
  of dumb sensors under the {SENMA} paradigm,'' \emph{{IEEE} Signal Process.
  Lett.}, vol.~12, no.~10, pp. 709--712, Oct. 2005.

\bibitem{spawc06}
S.~Marano, V.~Matta, and L.~Tong, ``Secrecy in cooperative {SENMA} with
  unauthorized intrusions,'' in \emph{Proceedings of IEEE Workshop on Signal
  Processing Advances in Wireless Communications (SPAWC)}, Cannes, France, July
  2-5 2006.

\bibitem{tong:C-SENMA-LDPC}
Z.~Yang and L.~Tong, ``On the error exponent and the use of {LDPC} codes for
  cooperative sensor networks with misinformed nodes,'' \emph{{IEEE} Trans.
  Inf. Theory}, vol.~53, no.~9, pp. 3265--3274, Sep. 2007.

\bibitem{boyd-infocom}
S.~S.~Boyd, A.~Ghosh, and B.~S.~D. Prabhakar, ``Gossip algorithms: Design,
  analysis and applications,'' in \emph{Proc.\ of INFOCOM}, Miami, USA, March,
  13-17 2005, pp. 1653--1664.

\bibitem{running-cons}
P.~Braca, S.~Marano, and V.~Matta, ``Enforcing consensus while monitoring the
  environment in wireless sensor networks,'' \emph{{IEEE} Trans. Signal
  Process.}, vol.~56, no.~7, pp. 3375--3380, 2008.

\bibitem{asymptotic-rc}
P.~Braca, S.~Marano, V.~Matta, and P.~Willett, ``Asymptotic optimality of
  running consensus in testing binary hypotheses,'' \emph{{IEEE} Trans. Signal
  Process.}, vol.~58, no.~2, pp. 814--825, 2010.

\bibitem{kar-moura-stsp}
S.~Kar and J.~M.~F. Moura, ``Convergence rate analysis of distributed gossip
  (linear parameter) estimation: {Fundamental} limits and tradeoffs,''
  \emph{{IEEE} J. Sel. Topics Signal Process.}, vol.~5, no.~4, pp. 674--690,
  Aug. 2011.

\bibitem{mouraetal2011}
D.~Bajovic, D.~Jakoveti\'c, J.~Xavier, B.~Sinopoli, and J.~M.~F. Moura,
  ``Distributed detection via {Gaussian} running consensus: {Large} deviations
  asymptotic analysis,'' \emph{{IEEE} Trans. Signal Process.}, vol.~59, no.~9,
  pp. 4381--4396, Sep. 2011.

\bibitem{mouraetal2012}
D.~Jakoveti\'c, J.~M.~F. Moura, and J.~Xavier, ``Distributed detection over
  noisy networks: {Large} deviations analysis,'' \emph{{IEEE} Trans. Signal
  Process.}, vol.~60, no.~8, pp. 4306--4320, Aug. 2012.

\bibitem{mouraetal2012_2}
D.~Bajovic, D.~Jakoveti\'c, J.~M.~F. Moura, J.~Xavier, and B.~Sinopoli, ``Large
  deviations performance of consensus+innovations distributed detection with
  {non-Gaussian} observations,'' \emph{{IEEE} Trans. Signal Process.}, vol.~60,
  no.~11, pp. 5987--6002, Nov. 2012.


\bibitem{CattivelliSayedDetection}
------, ``Distributed detection over adaptive networks using diffusion
  adaptation,'' \emph{{IEEE} Trans. Signal Process.}, vol.~59, no.~5, pp.
  1917--1932, 2011.



\bibitem{SayedSPmag}
A.~H. Sayed, S.-Y. Tu, J.~Chen, X.~Zhao, and Z.~J. Towfic, ``Diffusion
  strategies for adaptation and learning over networks,'' \emph{{IEEE} Signal
  Process. Mag.}, vol.~30, no.~3, pp. 155--171, 2013.

\bibitem{SayedNOW2014}
A.~H. Sayed, ``Adaptation, learning, and optimization over networks,'' in
  \emph{Foundations and Trends in Machine Learning}. Boston-Delft: NOW Publishers, 2014, vol.~7, no. 4--5, pp.
  311--801.

\bibitem{SayedprocIEEE}
------, ``Adaptive networks,'' \emph{Proc. IEEE}, vol. 102, no.~4, pp.
  460--497, Apr. 2014.

\bibitem{chen-sayed-IT1}
J.~Chen and A.~H. Sayed, ``On the learning behavior of adaptive
  networks---{Part I}: {Transient} analysis,'' \emph{{IEEE} Trans. Inf.
  Theory}, vol.~61, no.~6, pp. 3487--3517, Jun. 2015.

\bibitem{chen-sayed-IT2}
------, ``On the learning behavior of adaptive networks---{Part II}:
  {Performance} analysis,'' \emph{{IEEE} Trans. Inf. Theory}, vol.~61, no.~6,
  pp. 3518--3548, Jun. 2015.

\bibitem{TowficChenSayedIT2016}
Z.~Towfic, J.~Chen, and A.~H. Sayed, ``Excess-risk of distributed stochastic
  learners,'' \emph{{IEEE} Trans. Inf. Theory}, vol.~62, no.~10, pp.
  5753--5785, Oct. 2016.

\bibitem{BracaetalIT}
V.~Matta, P.~Braca, S.~Marano, and A.~H. Sayed, ``Diffusion-based adaptive
  distributed detection: Steady-state performance in the slow adaptation
  regime,'' \emph{{IEEE} Trans. Inf. Theory}, vol.~62, no.~8, pp. 4710--4732,
  Aug. 2016.

\bibitem{MattaSIPN16}
------, ``Distributed detection over adaptive networks: Refined asymptotics and
  the role of connectivity,'' \emph{IEEE Trans.\ Signal and Inf.\ Process.\
  over Networks}, vol.~2, no.~4, pp. 442--460, Dec. 2016.

\bibitem{MaranoSayedIT19}
S.~Marano and A.~H. Sayed, ``Detection under one-bit messaging over adaptive
  networks,'' \emph{{IEEE} Trans. Inf. Theory}, vol.~65, no.~10, pp.
  6519--6538, Oct. 2019.

\bibitem{ICASSP2020}
------, ``Adaptation and learning in multi-task decision systems,'' in
  \emph{Proc.\ of the 45th IEEE International Conference on Acoustics, Speech
  and Signal Processing ({ICASSP} 2020)}, Barcelona, May 4-8 2020.

\bibitem{maranosayedsubmitted2020}
------, ``Decision learning and adaptation over multi-task networks,''
  June 2020, \emph{submitted}.

\bibitem{ChenJSTSP17}
J.~Chen, C.~Richard, and A.~H. Sayed, ``Multitask diffusion adaptation over
  networks with common latent representations,'' \emph{{IEEE} J. Sel. Topics
  Signal Process.}, vol.~11, no.~3, pp. 563--579, Apr. 2017.







\bibitem{Sayed2008adaptive}
A.~H. Sayed, \emph{Adaptive Filters}.  NY: Wiley, 2008.

\bibitem{Hassibi-Sayed-Kailath} 
B.~Hassibi, A.~H.~Sayed and T.~Kailath, ``H$_{\infty}$ optimality of the LMS algorithm,'' \emph{{IEEE} Trans. Signal Process.,} vol.~44, no.~2, pp. 267-280, Feb.~1996.

\bibitem{basseville-book}
M.~Basseville and I.~V. Nikiforov, \emph{Detection of Abrupt Changes: Theory
  and Application}. Englewood Cliffs,
  N.J: Prentice-Hall, 1993.

\bibitem{tartakovsky-book}
A.~Tartakovsky, I.~Nikiforov and M.~Basseville, \emph{Sequential Analysis. Hypothesis
  Testing and Changepoint Detection}.
  Boca Raton, FL, USA: CRC Press, Taylor \& Francis Group, 2015.

\bibitem{poorbook}
H.~V. Poor, \emph{An Introduction to Signal Detection and Estimation}. New York: Springer-Verlag, 1988.

\bibitem{CT2}
T.~M. Cover and J.~A. Thomas, \emph{Elements of Information Theory},
  2nd~ed. New Jersey, USA:
  Wiley-Interscience, 2006.

\bibitem{wald}
A.~Wald, \emph{Sequential Analysis}. New York: Dover, 1947.

\bibitem{wald-wolfowitz-AMS}
A.~Wald and J.~Wolfowitz, ``Optimum character of the sequential probability
  ratio test,'' \emph{The Annals of Statistics}, vol.~19, no.~3, pp. 326--339,
  Sep. 1948.

\bibitem{Sayed-CooperativeSensing}
Q.~Zou, S.~Zheng and A.~H. Sayed, ``Cooperative sensing via sequential detection,'' \emph{{IEEE} Trans. Signal Process.}, vol.~58, no.~12, pp. 6266--6283, Dec. 2010.

\bibitem{afanaseva}
L.~G. Afanas'eva and E.~V. Bulinskaya, ``Certain asymptotic results for random
  walks in a strip,'' \emph{Theory of Probability and its Applications}, vol.
  XXIX, no.~4, pp. 677--693, 1985.
  
\bibitem{borovkov}
A.~A.~Borovkov ``On a walk in a strip with inhibitory boundaries,'' \emph{Institute of Mathematics, Siberian Branch, Academy of Sciences of the USSR.} Translated from Matematicheskie Zametki, vol.~17, no.~4, pp.~649-657, April 1975.

\bibitem{Page}
E.~Page, ``Continuous inspection schemes,'' \emph{Biometrika}, vol.~41, pp.
  100--115, Jan. 1954.

\bibitem{crowder1987}
S.~V. Crowder, ``A simple method for studying run-length distributions of
  exponentially weighted moving average charts,'' \emph{Technometrics},
  vol.~29, no.~4, pp. 401--407, 1987.

\bibitem{siegmund}
D.~Siegmund, \emph{Sequential Analysis: Tests and Confidence Intervals}. New York: Springer-Verlag, 1985.

\bibitem{SIR-1927}
W.~O. Kermack and A.~G. McKendrick, ``A contribution to the mathematical theory
  of epidemics,'' \emph{Proc. Roy. Soc. Lond. A}, vol. 115, pp. 700--721, 1927.

\bibitem{allen-2017-primer}
L.~J.~S. Allen, ``A primer on stochastic epidemic models: Formulation,
  numerical simulation, and analysis,'' \emph{Infectious Disease Modelling},
  vol.~2, pp. 128--142, 2017.
  
\bibitem{MAST-NC}
P.~Braca, D.~Gaglione, S.~Marano, L.~M.~Millefiori, P.~Willett, and K.~Pattipati, \emph{Quickest Detection of Critical COVID-19 Phases:
When Should Restrictive Measures Be Taken?}  Nov. 2020, \emph{submitted}. Accessible at http://arxiv.org/abs/2011.11540

\bibitem{MAST-SPL}
------ \emph{Quickest Detection of COVID-19 Pandemic Onset}. Nov. 2020, \emph{to be submitted}. Accessible at http://arxiv.org/abs/2011.10502




\bibitem{FellerBookV2}
W.~Feller, \emph{An Introduction to Probability and Its Applications,
  Volume~2}. New York: John Wiley \&  Sons, 1971.

\bibitem{ross-book}
S.~Ross, \emph{Stochastic Processes}, 2nd~ed. New York: John Wiley \& Sons, Inc., 1996.

\bibitem{Zorichbook2}
V.~A. Zorich, \emph{Mathematical Analysis {II}}. Berlin: Springer, 2004.

\bibitem{Zorichbook}
------, \emph{Mathematical Analysis {I}}. Berlin: Springer, 2004.

\bibitem{Avazzadeh}
Z.~Avazzadeh and M.~Heydari, ``Integral mean value method for solving
  {Fredholm} integral equations of the second kind on the half line,''
  \emph{Journal of Multidisciplinary Engineering Science Studies (JMESS)},
  vol.~2, no.~1, pp. 230--235, Jan. 2016.

\end{thebibliography}


\end{document}